\documentclass[ALICE,manyauthors]{cernphprep}

\usepackage[comma,square,numbers,sort&compress]{natbib}
\usepackage{hyperref}
\usepackage{lineno}
\usepackage{xspace}
\usepackage[usenames,dvipsnames]{color}

\newcommand{\pt}{\ensuremath{p_{\mathrm{T}}}\xspace}
\newcommand{\pT}{\ensuremath{p_{\mathrm{T}}}\xspace} 
\newcommand{\raa}{\ensuremath{R_{\mathrm{AA}}}\xspace}

\newcommand{\Npart}{\ensuremath{N_{\mathrm{part}}}\xspace}
\newcommand{\Ncoll}{\ensuremath{N_{\mathrm{coll}}}\xspace}
\newcommand{\avNpart}{\ensuremath{\langle N_\mathrm{part} \rangle}\xspace}
\newcommand{\avNcoll}{\ensuremath{\langle N_\mathrm{coll} \rangle}\xspace}
\newcommand{\avTAA}{\ensuremath{\langle T_\mathrm{AA} \rangle}\xspace}
\newcommand{\avTpPb}{\ensuremath{\langle T_\mathrm{pPb} \rangle}\xspace}
\newcommand{\rppb}{\ensuremath{R_{\mathrm{pPb}}}\xspace}
\newcommand{\snn}{\ensuremath{\sqrt{s_{\mathrm{NN}}}}\xspace}
\newcommand{\s}{\ensuremath{\sqrt{s}}\xspace}

\begin{document}%

\begin{titlepage}
\PHyear{2018}
\PHnumber{025}      
\PHdate{16 Feb}  
%

\title{Transverse momentum spectra and nuclear modification
    factors of charged particles in pp, p--Pb and Pb--Pb collisions at the LHC}
\ShortTitle{Charged-particle spectra and nuclear modification factors}   

\Collaboration{ALICE Collaboration\thanks{See Appendix~\ref{app:collab} for the list of collaboration members}}
\ShortAuthor{ALICE Collaboration} 

\begin{abstract}
We report the measured transverse momentum (\pt) spectra of primary charged
particles from pp, \mbox{p--Pb} and Pb--Pb collisions at a center-of-mass
energy \snn = 5.02 TeV in the kinematic range of $0.15<\pt<50$~
GeV/$c$ and $|\eta|< 0.8$.  A significant improvement of systematic uncertainties 
motivated the reanalysis of data in pp and Pb--Pb collisions at \snn = 2.76 TeV, as well as in p--Pb collisions at \snn = 5.02 TeV, 
which is also presented. Spectra from Pb--Pb collisions are presented
in nine centrality intervals and are compared to a reference
spectrum from pp collisions scaled by the number of binary
nucleon-nucleon collisions. For central collisions, the \pt
spectra are suppressed by more than a factor of 7 around 6--7~GeV/$c$
with a significant reduction in suppression towards higher momenta up
to 30 GeV/$c$.  
The nuclear modification factor \rppb, constructed from the pp and \mbox{p--Pb}
spectra measured at the same collision energy, is consistent with
unity above 8~GeV/$c$.
While the spectra in both pp and Pb--Pb collisions are substantially harder at  $\snn = 5.02$~TeV compared to 2.76~TeV, the nuclear modification factors show no significant collision energy dependence. The obtained results should provide further constraints on the parton energy loss calculations to determine the transport properties of the hot and dense QCD matter.

\end{abstract}
\end{titlepage}
\setcounter{page}{2}

%
\section{Introduction}

The properties of hot and dense deconfined QCD matter, the Quark-Gluon Plasma (QGP), which is formed in high-energy heavy-ion collisions, can be characterized by the measurement of high transverse momentum particles produced by hadronisation of hard scattered partons in the early stage of the collision. It is expected that these partons lose energy by interactions with the hot and dense QCD matter, which leads to jet quenching~\cite{Bjorken:1982tu}. Manifested also as a suppression of high-\pt particles, jet quenching enables the extraction of the properties of the deconfined medium, in particular its transport coefficient~$\hat{q}$~\cite{Burke:2014aaa}.

The modification of high-\pt particle production is quantified with the nuclear modification factor \raa, defined as the ratio of the charged-particle \pt spectrum in A--A collisions to that in pp collisions scaled by the average number of binary nucleon--nucleon collisions $\avNcoll$ for a given centrality class of A--A collisions,
\begin{equation}
\raa=  \frac{{\rm d}N^{{\rm AA}}/ {\rm d}\pT}{\avNcoll {\rm d}N^{{\rm pp}}/{\rm d}\pT} =  \frac{{\rm d}N^{{\rm AA}}/{\rm d}\pT}{\avTAA {\rm d}\sigma^{{\rm pp}}/{\rm d}\pT},
\end{equation}  
where  $N^{{\rm AA}}$ and $N^{{\rm pp}}$ are the charged-particle yields in A--A and pp collisions and $\sigma^{{\rm pp}}$ is the production cross section in pp collisions, respectively. The average nuclear overlap function, $\avTAA=\avNcoll/ \sigma^{\rm NN}_{\rm inel}$, which depends on the collision centrality, is determined from the Glauber model of the nuclear collision geometry  \cite{Loizides:2017aya}, 
 where $\sigma^{\rm NN}_{\rm inel}$ is the total inelastic nucleon-nucleon cross section. 
Over the years, a number of results on \raa have been reported by experiments at the Relativistic Heavy-Ion Collider (RHIC) and at the Large Hadron Collider (LHC).  At RHIC, the yields of charged hadrons \cite{Adcox:2001jp,Adler:2002xw} or neutral pions \cite{Adare:2008akl} 
measured in the central \mbox{Au--Au} collisions at $\snn= 130$ and 200~GeV were found to be suppressed by a factor of about 5 in the \pt range of 5--25~GeV/$c$, indicating for the first time strong medium effects on hadron production. The first \raa measurements for charged particles at the LHC  \cite{Aamodt:2010jd, Abelev:2012hxa, CMS:2012aa, Aad:2015aeg} have shown that in central Pb--Pb collisions at $\snn = 2.76$~TeV the yields are suppressed by a factor of up to 7 for $\pt=6$--7~GeV/$c$. For larger \pt, the suppression decreases, but remains significant (a factor of about 2) in the range of 30--150~GeV/$c$.

The first Pb--Pb collisions at $\snn=5.02$~TeV were delivered by the LHC in 2015. Data in pp collisions at the same energy were also collected by the LHC experiments, allowing for a direct comparison of particle production in pp, p--Pb and Pb--Pb collisions. The first results on charged-particle \raa at $\snn = 5.02$~TeV have recently become available from the CMS Collaboration \cite{Khachatryan:2016odn}, showing that in central Pb--Pb collisions charged-particle production is suppressed by a factor of 7--8 for $\pt=6$--9~GeV/$c$. The suppression 
continues up to the highest \pt measured and approaches unity in the vicinity of \pt = 200~GeV/$c$.

Measurements of p--Pb collisions at the LHC were performed to establish whether the initial state of the colliding nuclei plays a role in the observed suppression of high-\pt hadron production in Pb--Pb collisions. 
The \rppb was found to be consistent with unity for \pt up to a few tens of GeV/$c$, indicating that in this domain initial state effects do not influence particle production \cite{Abelev:2012mj,Abelev:2014dsa,Khachatryan:2015xaa,Aad:2016agh}. 

In this paper, we report the measurement of transverse momentum spectra of charged particles in pp and Pb--Pb collisions at $\snn=5.02$~TeV. The resulting \pt spectra are used to determine the nuclear modification factors in Pb--Pb collisions at the highest energy currently accessible at the LHC. The \pt spectrum measured in pp collisions at the same collision energy as p--Pb is also used as the reference to calculate \rppb. These measurements allow us to compare the particle production in pp, p--Pb and Pb--Pb collisions at the same \snn, for the first time with ALICE at the LHC. In addition, we report a reanalysis of data collected in pp and Pb--Pb collisions at $\snn=2.76$~TeV, and in p--Pb collisions at $\snn=5.02$~TeV. The reanalysis is warranted by significant improvements in track selection and efficiency corrections, which benefit from the experience accumulated in the past years as well as better knowledge of the particle production at the LHC energies. This leads to significantly-reduced systematic uncertainties by a factor of about 2 as compared to previously published results \cite{Abelev:2012hxa,Abelev:2013ala,Abelev:2014dsa}, which the current analysis supersedes. The increase in \snn from 2.76~TeV to 5.02~TeV for Pb--Pb collisions leads to $\sim20\%$ increase in the particle multiplicity \cite{Adam:2015ptt} indicating that the larger medium density is reached at the higher collision energy. We characterize this medium by comparing the \pt spectra and nuclear modification factors measured at the two energies. 
\section{Experiment and data analysis}

The data in Pb--Pb and pp collisions at \snn = 2.76~TeV and in p--Pb collisions at \snn = 5.02~TeV were collected with the ALICE apparatus \cite{Aamodt:2008zz} in 2010, 2011 and 2013, respectively. Details on the ALICE experimental conditions and the detector performance are given in \cite{Abelev:2014ffa}.  The data in Pb--Pb and pp collisions at \snn = 5.02~TeV were recorded in 2015.

\subsection{Trigger and event selection}
The analysis is based on tracking information from the Inner Tracking
System (ITS) \cite{Dellacasa:1999out} and the Time Projection Chamber (TPC) \cite{Dellacasa:2000rtu}, both are located in
the central barrel of the experiment and embedded in a solenoidal
magnetic field of 0.5 T parallel to its axis.

The minimum-bias (MB) interaction trigger was based on signals from
the forward scintillator arrays (V0A and V0C) and the two innermost
layers of the ITS, the Silicon Pixel Detector (SPD), in coincidence
with two beam bunches crossing in the ALICE interaction region.  The
pp collisions at \s~=~2.76 TeV were selected requiring a signal in
either one of the V0A or the V0C detectors or in the SPD.  The Pb--Pb collisions at
\snn~=~2.76 TeV were selected based on different combinations of hits in the SPD and either V0A
or V0C detector. The efficiency for hadronic interactions is  approximately 100\% in the 0--80\% centrality range considered in this analysis, see details in \cite{Abelev:2014ffa}. For measurements of pp, p--Pb and Pb--Pb collisions at \snn = 5.02~TeV the trigger required a signal in both V0A and V0C
detectors.

The offline event selection was optimized to reject beam-induced
background in all collision systems. The background events were efficiently
rejected by exploiting the timing signals in the two V0 detectors. In Pb-Pb
collisions background was also rejected exploiting the correlation between
the arrival times measured in each neutron Zero Degree Calorimeter (ZDC), positioned
on both sides of the interaction point at 114.0~m for pp and Pb--Pb data at \snn~=~2.76 TeV and at 112.5~m for the rest data sets. The contamination
from electromagnetic interactions in Pb--Pb collisions was strongly
suppressed using signals from the ZDCs (see \cite{Abelev:2014ffa} for details).

The primary event vertex is determined with tracks from the central
barrel. For the analysis of pp collisions, if no vertex is found using
tracks, the vertex reconstruction is performed using SPD tracklets; track segments reconstructed based on the information from the two innermost layers of the ITS. To ensure a uniform acceptance and reconstruction efficiency in the
pseudorapidity region $|\eta|< 0.8$, only events with a reconstructed
vertex within $\pm$10~cm from the center of the detector along the
beam direction are used. It corresponds to around 2 standard deviations from the mean of the interaction region 
distribution (Gaussian shape) determined for all collisions systems and energies.

In Pb--Pb collisions, the centrality quantifies the fraction of the
geometrical cross-section of the colliding nuclei, and it is related
to their geometrical overlap region.  It is determined using the sum
of the amplitudes of the V0A and V0C signals \cite{ALICE-PUBLIC-2018-011}.
The analysis is limited to the 0--80\% most central events, to ensure
that effects of trigger inefficiency and contamination by
electromagnetic processes \cite{ALICE:2012aay}, as well as possible biases
in the selection of more peripheral events \cite{Morsch:2017toi}, are negligible. The average
quantities characterizing a centrality class, such as the mean number of
participants \avNpart, the mean number of binary collisions \avNcoll or the
average nuclear overlap function \avTAA were obtained~\cite{ALICE-PUBLIC-2018-011} by fitting the experimental
distributions with a Glauber Monte Carlo model \cite{Loizides:2017aya}, coupled to the model of particle production with $f \cdot \Npart + (1 - f) \cdot \Ncoll$ particle sources, each source producing particles according to a Negative Binomial Distribution (NBD). This approach is inspired by two-component models \cite{Kharzeev:2004if,Deng:2010mv}, which decompose nucleus-nucleus collisions into soft and hard interactions, where the soft interactions produce particles with an average multiplicity proportional to \Npart, and the probability for hard interactions to occur is proportional to \Ncoll. The fit parameter $f$ represents the contribution of soft processes to the particle production and amounts to about $0.8$ for the two energies. In this calculations, we used an inelastic nucleon--nucleon cross-section $\sigma_{{\rm NN}}= (67.6 \pm 0.6)$~mb for \snn = 5.02~TeV and $\sigma_{{\rm NN}} =
(61.8 \pm 0.9)$~mb for \snn = 2.76~TeV, obtained by interpolation \cite{Loizides:2017aya} of the existing world data.  

In p--Pb collisions, the average quantities \avNpart, \avNcoll and  \avTpPb were determined \cite{ALICE-PUBLIC-2018-011} following the procedure described in \cite{Adam:2015ary}, with the updated inelastic nucleon--nucleon cross-section $\sigma_{{\rm NN}}= (67.6 \pm 0.6)$~mb at \snn = 5.02~TeV and nuclear density function. In order to omit potential biases on the \pt spectra related to p--Pb collision centrality determination \cite{Adam:2015ary}, only p--Pb events in the 0--100\% centrality interval were used in the presented analysis. 

The number of events satisfying the trigger and offline selection criteria for various collision systems and energies are listed in Table~\ref{tab:data}.
\begin{table}[t]
\centering
\begin{tabular}{@{} c|c|c @{}} 
collision system  & \snn = 2.76 TeV & \snn = 5.02 TeV  \\
\hline
pp     &  52 M  & 109 M  \\
p--Pb & - & 107 M \\
Pb--Pb (0--80\%)& 13 M  & 20 M \\
\hline
\end{tabular}
\caption{\label{tab:data} Number of events used in the analysis for various systems and energies. The analysis of Pb--Pb events was performed for the 0--80\% centrality range.}
\end{table}

\subsection{Track selection}
Primary charged particles are measured in the kinematic range $|\eta|<0.8$ and $0.15 < \pt < 50$~GeV/$c$.  
A primary charged particle is defined \cite{Acharya:2017rih} to be a charged particle with a mean proper lifetime $\tau$ larger than 1~cm/$c$ which is either produced directly in the interaction, or from decays of particles with $\tau$ smaller than 1~cm/$c$,
excluding particles produced in interactions with the detector material.
The track-selection criteria were identical for all data sets and were optimized for best track quality and minimal contamination from secondary particles.
Each track is required to have:

\begin{itemize}
\item at least 2 hits in the ITS detector, of which at least one hit is in the two innermost (SPD) layers;
\item the length $L$ (in cm) of its projection curve calculated in the TPC readout plane, excluding the information from the pads at the sector boundaries ($\sim3$~cm from the sector edges), larger than $A - B \cdot \pt^{C}$, with $A=130$~cm, $B=1.0~{\rm cm} \cdot ({\rm GeV}/c)^{-C}$, $C=-1.5$ and \pt in units of GeV/$c$;
\item the number of crossed TPC pad rows larger than $0.85 \cdot L$ (the height of pad rows varies from 7.5~mm to 15~mm \cite{Dellacasa:2000rtu}); a TPC readout pad row is considered crossed if there is a cluster in this row and in any of its neighboring 2 rows;
\item the number of TPC clusters (one cluster per pad row) larger than $0.7 \cdot L$; 
\item the ratio of crossed TPC pad rows to the number of findable TPC clusters (maximum number of clusters which can be assigned to a track in the TPC fiducial volume, excluding the information from the pads at the sector boundaries) larger than 0.8;
\item the fraction of TPC clusters shared with another track lower than 0.4;
\item the fit quality for the ITS and TPC track points satisfying $\chi_{{\rm ITS}}^{2}/{N_{\rm hits}}<36$ and $\chi_{{\rm TPC}}^{2}/{N_{\rm clusters}}<4$, respectively;
\item $\chi^{2}_{\rm TPC-ITS}<36$, where $\chi^{2}_{\rm TPC-ITS}$ is calculated comparing the track parameters of the helix fit from the combined ITS+TPC track reconstruction to that derived only from the TPC and constrained by the interaction point, see details in \cite{Abelev:2012hxa};
\item the distance of closest approach to the primary vertex in the transverse plane $|{\rm DCA}_{{\rm xy}}|<A+B\cdot \pt^{C}$, with $A=0.0182$~cm, $B=0.035~{\rm cm} \cdot ({\rm GeV}/c)^{-C}$, $C=-1.0$ and \pt in units of GeV/$c$;  and along the beam axis $|{\rm DCA}_{{\rm z}}|<2$~cm.
\end{itemize}

\subsection{Corrections}

The data are presented as differential cross sections for inelastic (INEL) pp collisions 
\begin{equation} \label{eq:cross_sec}
   \frac{{\rm d}^{2} \sigma} {{\rm d}\eta {\rm d}\pt}
  =  \sigma_{\rm {MB}}^{\rm {pp}} \cdot \frac{1}{N_{{\rm ev}}^{{\rm MB}}} \frac{{\rm d}^{2} N} {{\rm d}\eta {\rm d}\pt} 
  \equiv  \sigma_{\rm {MB}}^{\rm {pp}} \cdot \frac{N^{\rm {rec}}(\Delta\eta,\Delta \pt) \cdot C(\Delta\eta, \Delta \pt)}{N_{\rm {ev}}^{{\rm rec}} \cdot \Delta\eta \Delta\pt} \cdot \epsilon_{\rm {VZ}}, 
\end{equation}
and transverse momentum spectra for non-single diffractive (NSD) p--Pb and centrality-selected INEL Pb--Pb collisions
\begin{equation} \label{eq:inv_yields}
 \frac{1}{N_{{\rm ev}}}  \frac{{\rm d}^{2} N} {{\rm d}\eta {\rm d}\pt}
  \equiv  \frac{N^{\rm {rec}}(\Delta\eta, \Delta \pt) \cdot C(\Delta\eta,\Delta \pt)}{N_{\rm {ev}}^{{\rm rec}} \cdot \Delta\eta \Delta\pt} \cdot \epsilon_{\rm {MB}}  \cdot \epsilon_{\rm {VZ}},
\end{equation}
which are obtained by correcting the charged particle yields $N^{\rm {rec}}$ reconstructed in the $(\Delta\eta,\Delta \pt)$ intervals for all detector
effects that either influence the event reconstruction, and thus are relevant only for the overall normalization (event-level corrections), or influence the track reconstruction and are relevant for both the spectral shape and normalization (track-level corrections). The $\epsilon_{{\rm MB}}$ and  $\epsilon_{{\rm  VZ}}$ denote the MB trigger and event vertex reconstruction efficiencies, and $C(\Delta\eta, \Delta \pt)$ are track-level correction factors. One should note that the $\epsilon_{{\rm  VZ}}$ is calculated for the triggered events. In general, both the $\epsilon_{{\rm MB}}$ and  $\epsilon_{{\rm  VZ}}$ are multiplicity dependent. Details of the correction procedure and variables are described in the following. 

\subsubsection{Event-level corrections}
In Eq. \ref{eq:cross_sec} the minimum-bias cross section $\sigma_{\rm MB}^{\rm pp}$ in triggered pp collisions is determined by the van-der-Meer scans and depends on the trigger settings, it was measured to be $55.4 \pm 1.0$~mb at \s = 2.76~TeV \cite{Abelev:2013adr} and $51.2 \pm 1.2$~mb at  \s = 5.02~TeV \cite{Adam:2016rth}, with the MB trigger OR (V0A or V0C or SPD) and AND (V0A and V0C), respectively. The differential charged-particle yields ${\rm d}^{2} N/{\rm d}\eta {\rm d}\pt$ were calculated for the MB event class $\left({N_{\rm{ev}}^{\rm {MB}}}\right)$ by normalizing to the number of reconstructed events ${N_{\rm{ev}}^{\rm {rec}}}$, which have a reconstructed event vertex within $\pm$10~cm from of the center of the detector and correcting for the event vertex reconstruction efficiency $\epsilon_{\rm{VZ}}$. 

For INEL pp collisions, the $\epsilon_{\rm{VZ}}$ was estimated using the PYTHIA 8 (Monash 2013 tune) event generator
\cite{Sjostrand:2008aro, Skands:2014mon} and GEANT3 \cite{GEANT3} detector response model. The resulting values $\epsilon_{\rm{VZ}}= 88.3\% (97.7\%)$ at $\s = 2.76$ (5.02)~TeV were used for corrections.

For NSD p--Pb collisions, the efficiency of the trigger ($\epsilon_{\rm MB}$) and event vertex reconstruction ($\epsilon_{\rm VZ}$), as in Eq.~\ref{eq:inv_yields}, were estimated using GEANT3 detector simulation with a combination of event generators as described in \cite{Abelev:2012mj}. The obtained values $\epsilon_{\rm MB} =  99.2\%$ and $\epsilon_{\rm VZ}= 98.6\%$  were used for corrections.

For Pb--Pb collisions, the trigger and event vertex reconstruction is
fully efficient for the centrality intervals considered in this work, as estimated
using Monte Carlo simulations with GEANT3 and HIJING \cite{Wang:1991hta} as event
generator.

\begin{figure}[!hbt]
	\centering  
	\includegraphics[width=0.49\textwidth]{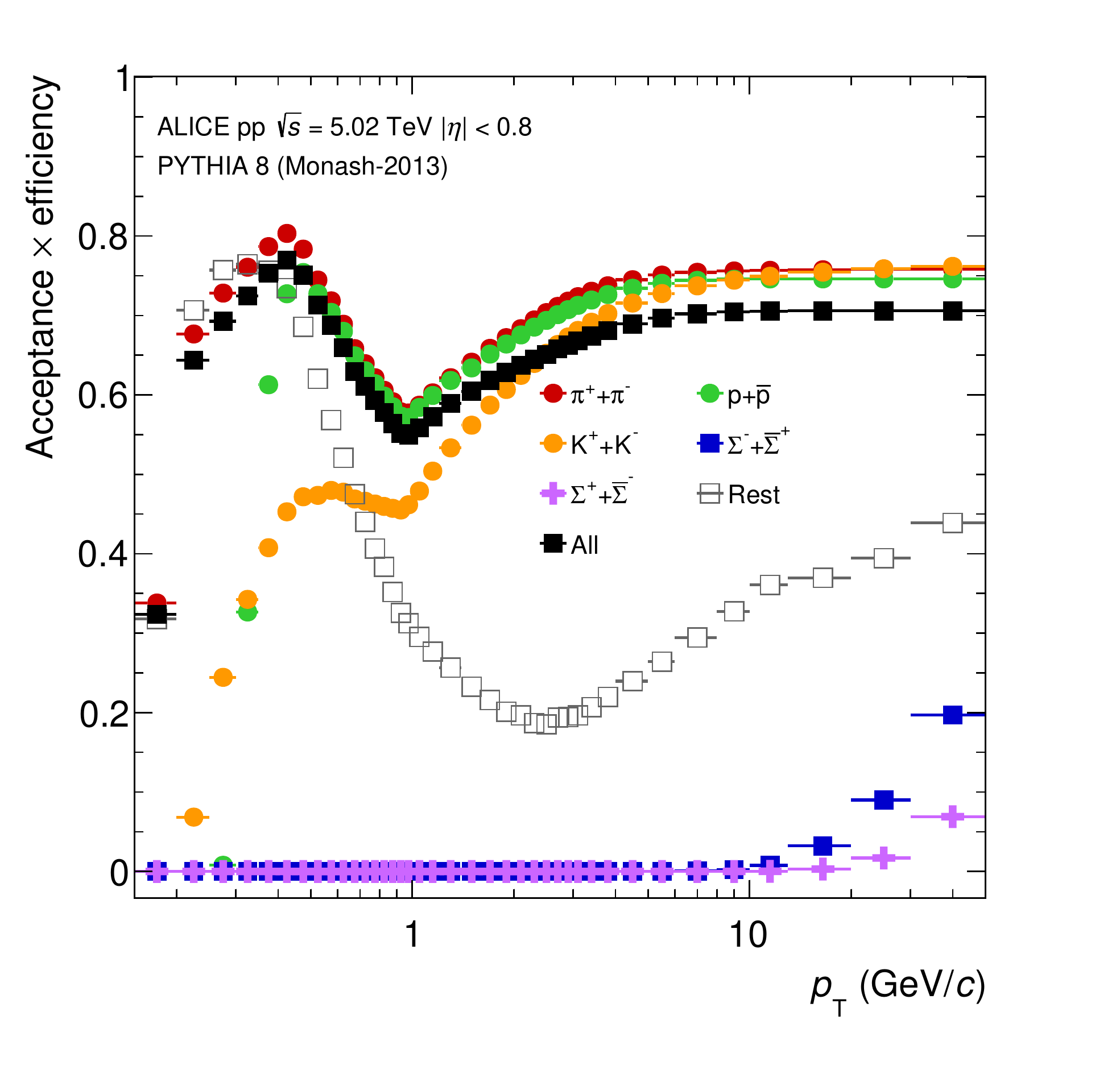}
	\includegraphics[width=0.49\textwidth]{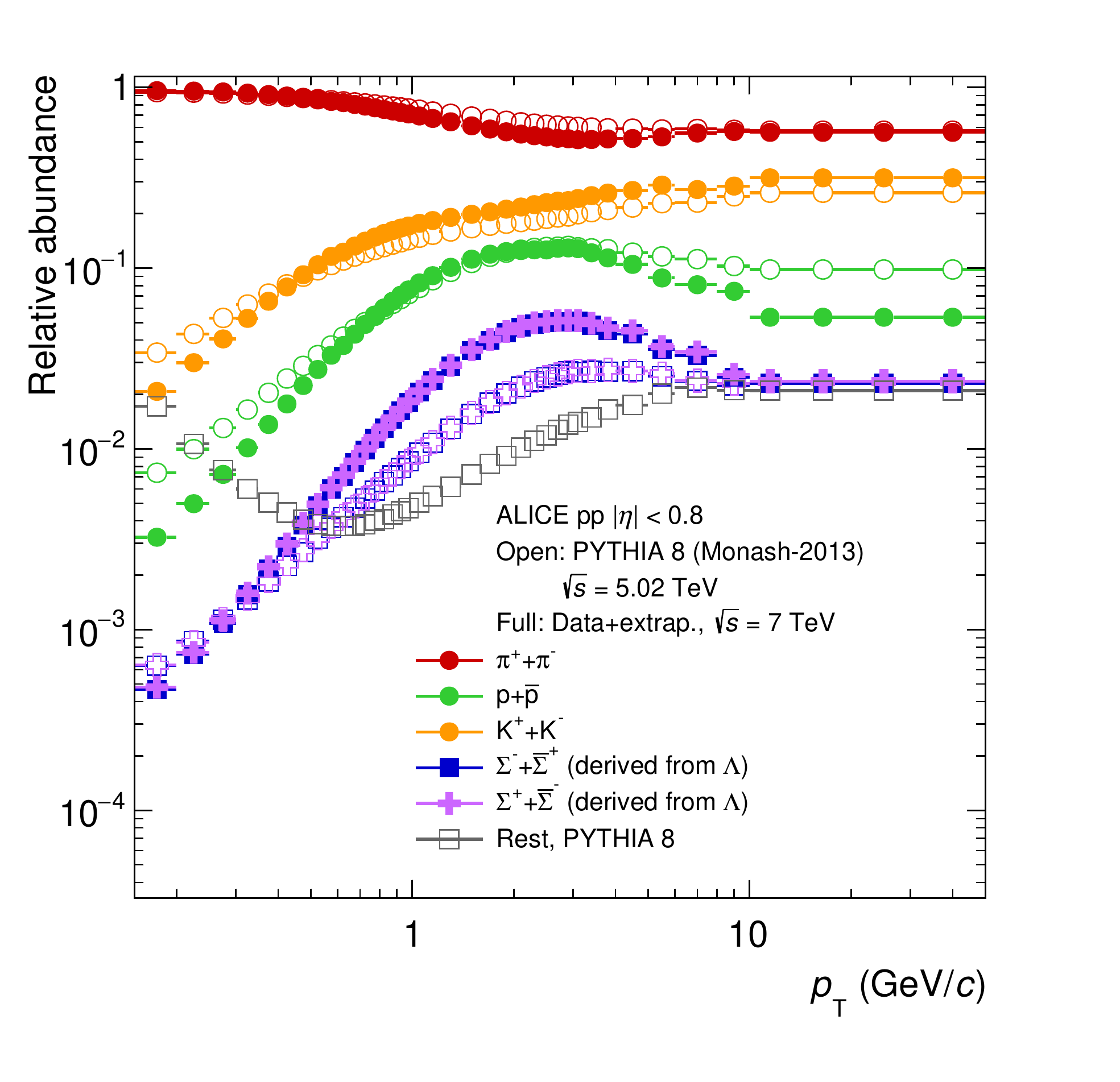}
    \caption{Left: Combined tracking efficiency and acceptance as a
      function of \pT for different particle species and the sum of all,
      obtained in Monte Carlo simulations of pp collisions at $\s =
      5.02$~TeV with PYTHIA 8 (Monash 2013 tune).  For $\pt > 1$~GeV/$c$ parameterizations are shown. The relative systematic uncertainties on parameterizations  are small ($<0.2\%$) and are not shown. The statistical uncertainties for $\pt < 1$~GeV/$c$ are smaller than the symbol size ($<0.5\%$).  Right: The relative particle abundances as a function of \pt in Monte Carlo (open symbols,  for $\s = 5.02$~TeV) and in data (full symbols, for $\s = 7$~TeV)
\cite{Adam:2015qaa,  Adam:2016dau, Adam:2017nat}. The data are extrapolated beyond the range of the measurements (see description in the text). The statistical and systematic uncertainties (combined $<1.6\%$) are not shown.}
	\label{fig:ParComp}
\end{figure}

\subsubsection{Track-level corrections}
The differential charged-particle yields ${\rm
  d}^{2} N/{\rm d}\eta {\rm d}\pt$ (Eqs \ref{eq:cross_sec} and \ref{eq:inv_yields}) are obtained from the reconstructed
yields of tracks $N^{\rm{rec}}(\Delta \eta, \Delta \pt)$ corrected using correction factors $C(\Delta\eta, \Delta \pt)$, which are products of 
acceptance, efficiency, purity and \pt resolution.

The efficiency and purity of the primary charged particle
reconstruction as well as acceptance correction for the pp, p--Pb and Pb--Pb data are calculated using Monte Carlo event generators with
GEANT3 detector modeling combined with data-driven corrections, which are discussed in detail in the following sections.

\paragraph{Tracking efficiency}
The efficiency of the primary charged particle
reconstruction is shown in Fig.~\ref{fig:ParComp} (left). While the low efficiency at low \pt is related to the strong track curvature caused by the 
magnetic field and to the energy loss in the detector material, the characteristic shape around \pt of
1 GeV/$c$ is caused primarily by the track length requirement. 
Tracks in this momentum range are more likely to cross the TPC sector boundaries 
and are thus reconstructed with lower efficiencies. The asymptotic value reached at high \pt reflects the acceptance
limitations (detector boundaries and active channels) of the measurement. 

The tracking efficiency depends on particle species, as can be seen in Fig.~\ref{fig:ParComp} (left), and was calculated using a detector simulation with the PYTHIA 8 (Monash 2013 tune) event generator and the GEANT3 transport code.
The efficiency is particularly species-dependent at low  $\pT$ (below 0.5 GeV/$c$) due to differences in ionization energy loss in the detector material, hadronic interaction cross-section or decay probability. 

A particular case is that of charged hyperons, for which the reconstruction
efficiency is very low and essentially negligible below 10~GeV/$c$, due to the fact that they decay before any significant interaction with the detector. For higher \pt, they reach the detector and can be observed with increasing efficiency. One should note that the
reconstruction efficiency is different for the $\Sigma^{+}$ and
$\Sigma^{-}$ hyperons in the \pt range considered, because of their different lifetimes. The tracking efficiency for other primary charged particle species,
including electrons, muons and $\Xi$ and $\Omega$ hyperons
(denoted as "Rest") is also shown.

In order to reduce statistical
fluctuations at high \pt, we parameterized the efficiency above
$\pt = 1$~GeV/$c$ for each particle species.  Each parameterization is a
combination of the universal (independent of particle species)
function $f(\pt) = a\left(1-b \cdot e^{-c \pt} \right)$ and the survival
probability $P(\pt) = e^{-d \cdot m/\pt \cdot \tau}$ that a particle
with the mass $m$ and a mean proper lifetime $\tau$ survives a minimal
distance $d$ before decaying. The fitting parameters ($a$, $b$ and $c$) are
determined from the fit to the efficiency calculated as an average of
efficiencies for stable particles. The calculations were
performed for $d=200$~cm, corresponding to the minimum track length in
the ITS and the TPC required in the analysis. 

The parametrized efficiencies shown in Fig.~\ref{fig:ParComp} (left) were used to 
determine data-driven correction factors in the efficiency rewieghting procedure, which is discussed below. 

\begin{figure}[!htb]
\centering 
\resizebox*{.49\columnwidth}{!}{\includegraphics{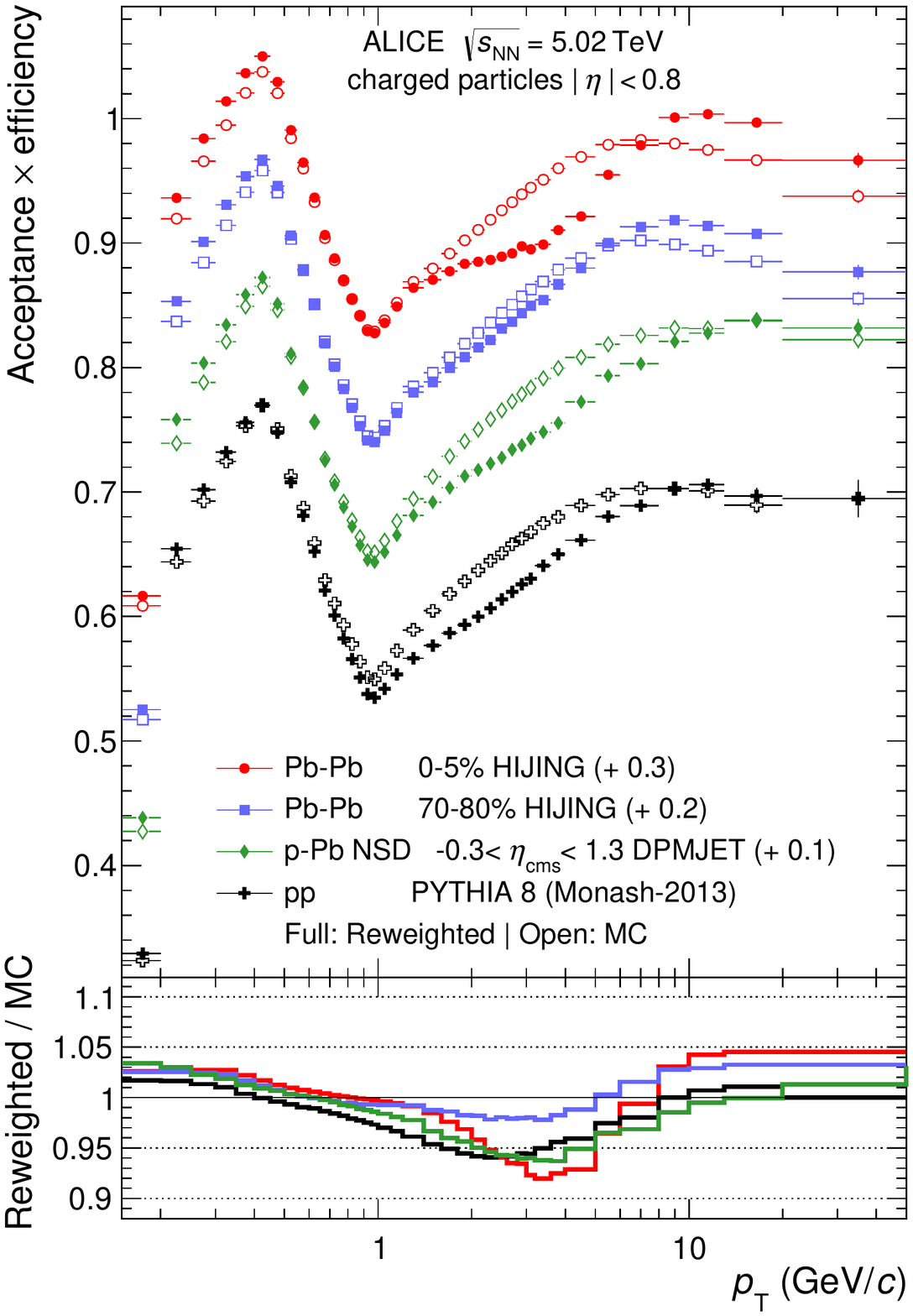}} 
\resizebox*{.49\columnwidth}{!}{\includegraphics{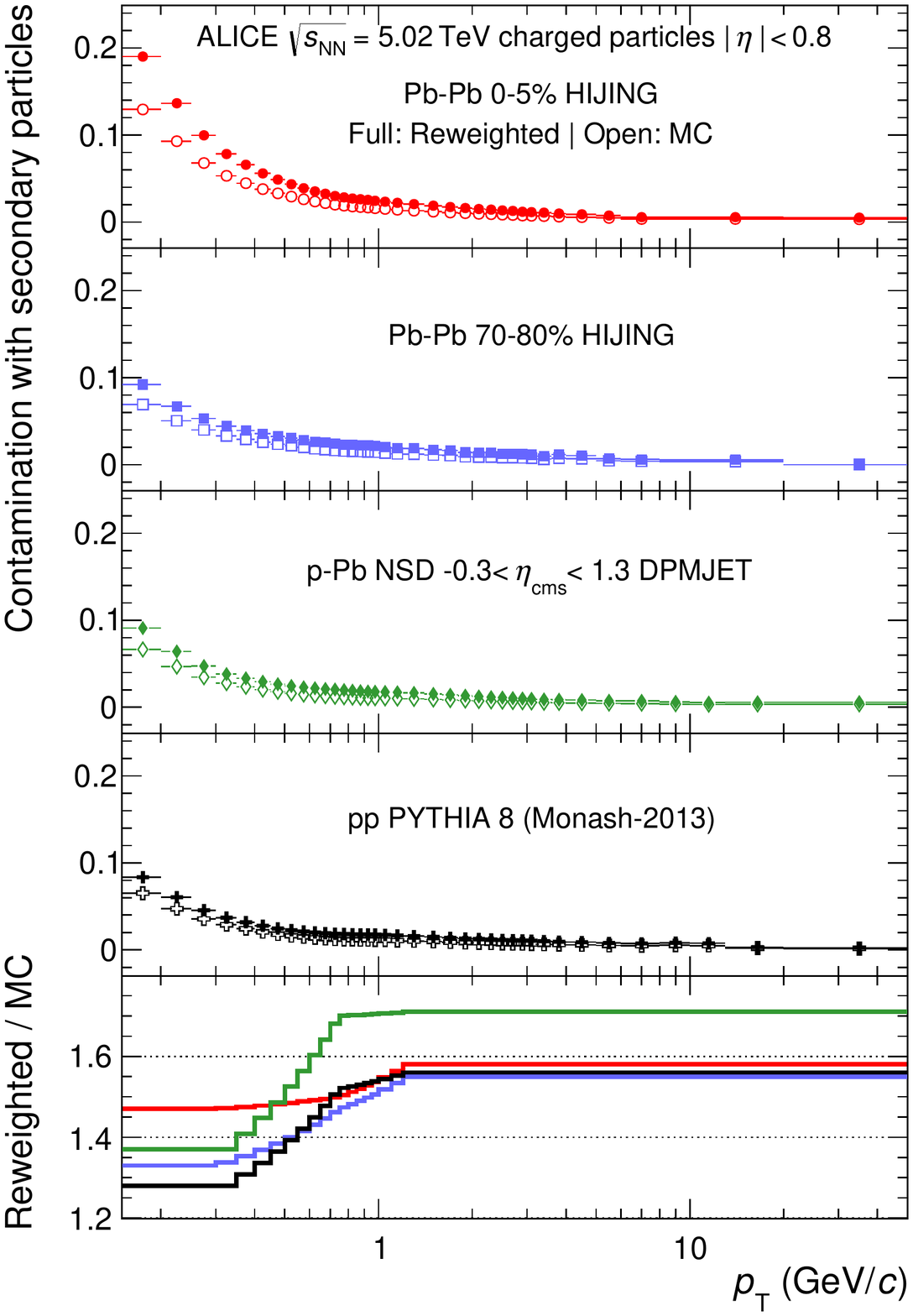}}
\caption{Left: Combined tracking efficiency and
    acceptance as a function of \pT for pp, p--Pb, central (0-5\%) and
    peripheral (70--80\%) Pb--Pb collisions determined using Monte Carlo simulations and a reweighting method (see text for details). For better visibility, the curves for p--Pb and Pb--Pb are offset by the indicated
    values.  The effect of the reweighting on the efficiency corrections is shown in the bottom panel. The systematic uncertainties of the reweighting ($<2.4\%$) are not shown.
    Right: Contamination from secondary particles estimated
    from Monte Carlo simulations and from the impact parameter fits in
    data (see text for details). The effect of the reweighting of secondary particles is shown in the bottom panel. The systematic uncertainties on the scaling factors ($<20\%$) are not shown.  
    \label{fig:eff}}
\end{figure}

\paragraph{Reweighting with measured particle composition}

The experimental knowledge of the primary particle composition has significantly improved recently at the LHC \cite{Adam:2015qaa, Adam:2016dau, Adam:2017nat,  Abelev:2014qqa, Adamova:2017elh, Abelev:2013haa, Abelev:2013vea, Abelev:2013xaa, Abelev:2014laa, Adam:2016tre}, which allows for a precise determination of the tracking efficiency. 
For the first time, we determined the tracking efficiency 
by reweighting the primary particle composition based on data driven method. 

In the right panel of Fig.~\ref{fig:ParComp}, the relative particle
abundances measured by ALICE in pp collisions at $\s=7$~TeV are
compared to those from Monte Carlo simulations with the PYTHIA 8
(Monash 2013 tune) event generator for $\s=5.02$~TeV. Charged pions,
kaons and protons were measured from \pt = 0.15, 0.2, and 0.3
GeV/$c$ to 20 GeV/$c$ \cite{Adam:2015qaa, Adam:2016dau}, respectively. It is known that Monte Carlo event
generators underestimate hyperon production substantially 
\cite{Abelev:2014qqa,Adamova:2017elh}. In particular, the $\Sigma^{+}(1385)$
and $\Sigma^{-}(1385)$ hyperons and their antiparticles are underestimated by a factor of 2--3 in the recent PYTHIA 8 tunes. 
The \pt spectra of $\Sigma^{\pm}$ hyperons have not been measured. 
Therefore, the \pt spectra of $\Sigma^{\pm}$ are approximated 
using the measured spectrum of $\Lambda$ hyperons \cite{Adam:2017nat}
scaled by the ratio of $\Sigma^{\pm}$
to $\Lambda$ hyperons from the Monte Carlo generator.  

Relative particle abundances measured in pp collisions at $\s=7$~TeV  
are used to reweight the tracking efficiency determined for $\s=2.76$ and 5.02~TeV collision energies, based on the experimental
knowledge that their energy dependence is weak \cite{Adam:2017nat}.
The relative abundance of other particle species containing electrons,
muons and $\Xi$ and $\Omega$ hyperons (denoted ``Rest" in
Fig.~\ref{fig:ParComp}) is taken from simulations without further
modification and has only a small influence on the final result ($<1\%$).
The measured \pt spectra of kaons, protons and $\Lambda$ are
extrapolated down to $\pt = 0.15$~GeV/$c$ using a parameterization
proposed by Bylinkin and Rostovtsev \cite{Bylinkin:2012eru}. 
For high \pt, beyond the reach of the identified particle measurement, the
relative abundances are assumed to be independent of \pt, as motivated
by pQCD \cite{deFlorian:2017lwf}.

The reweighting of the efficiency has also been applied in the
analysis of p--Pb and Pb--Pb data.  The relative particle abundances
obtained from Monte Carlo simulations with DPMJET (p--Pb)
\cite{Roesler:2000he} and HIJING (Pb--Pb) event generators are
reweighted using ALICE measurements of identified particle species
(pions, kaons, protons and $\Lambda$ hyperons) for p--Pb collisions at
\snn = 5~TeV \cite{Abelev:2013haa, Adam:2016dau} and Pb--Pb collisions
at \snn = 2.76~TeV
\cite{Abelev:2013vea,Abelev:2013xaa,Abelev:2014laa,Adam:2016tre}.  The
relative particle abundances at low \pt are determined by
extrapolating the measured \pt spectra of kaons, protons and $\Lambda$
hyperons down to $\pt = 0.15$~GeV/$c$ using a blast-wave
parameterization \cite{Schnedermann:1993art}. As in the pp case, for \pt
beyond the reach of these measurements, the relative abundances are
assumed to be independent of \pt.

In the left panel of Fig.~\ref{fig:eff} the combined tracking
efficiency and acceptance obtained from MC simulations (open symbols) and after reweighting (full symbols) is shown as a function of \pt for pp, p--Pb,
and central (0-5\%) and peripheral (70-80\%) Pb--Pb
collisions. The effect of the reweighting on the efficiency
corrections is shown in the bottom panel. It amounts to a difference
of about 7\% at \pT around 3 GeV/c for the most central Pb--Pb
collisions, and is lower in peripheral Pb--Pb collisions, p--Pb and pp
collisions. When comparing central to peripheral
Pb--Pb collisions, the importance of an increasing radial flow that
shifts the heavy $\Sigma^{\pm}$ baryons to larger momenta becomes
apparent.

\paragraph{Purity}
The contribution from secondary particles, i.e. products of weak
decays of kaons, $\Lambda$ hyperons and muons, and particles arising from
interactions in the detector material, was estimated using the transverse impact
parameter $d_{xy}$ distributions of particles in data and Monte Carlo
simulations.  Exploiting the differences,
especially in the tails, of the $d_{xy}$ distributions between primary
and secondary particles, the measured distributions were fitted by a
linear combination of $d_{xy}$ distributions (templates) for
primary and secondary particles obtained from the Monte Carlo simulations in
different $\pT$ bins (as described in more detail in
\cite{Abelev:2013vea}).  The effect of this data-driven correction,
shown in the bottom panel of Fig.~\ref{fig:eff} (right), depends on \pT
and is different for pp, p--Pb and Pb--Pb collisions. The resulting
contamination with secondary particles, i.e. the fraction of secondary particles in the sample of selected particles, ranges from 8.5\% in pp to 20\% in central Pb--Pb
collisions at $\pT = 0.15$~GeV/$c$ and decreases to around 1.0\% for
$\pT> 5$~GeV/$c$, as shown in the upper panel of Fig.~\ref{fig:eff} (right).

\begin{table} [!hpt]
                \centering 
                \begin{tabular}{|l |l |l |l|l|l|}
                \hline
                \textbf{Source of Uncertainty}  & \multicolumn{2}{c|} { \textbf{pp} } & \textbf{p--Pb} & \multicolumn{2}{c|} {\textbf{Pb--Pb}} \\ 
                 & 2.76 TeV & 5.02 TeV & 5.02 TeV &  2.76 TeV & 5.02 TeV \\ 
                \hline \hline
                Event selection                    & 0.9          & 0.5         & 0.1         & 1.5        & 0.14 \\
                Track selection                    & 0.4--3.8     & 0.6--3.5    &0.6--3.8     & 1.0--2.0   & 0.6--4.9 \\
                Secondary particles                & 0.5--5.1     & 0.0--2.8    & 0.0--2.1    & 0.0--4.0   & 0.0--4.5 \\
                Particle composition               & 0.1--1.6     & 0.2--2.4    & 0.4--2.2    & 0.0--2.0   & 0.2--2.0 \\
                Matching efficiency                & 1.0--4.0     & 0.0--1.1    & 0.3--3.2    & 0.2--2.0   & 0.2--1.2 \\
                Trigger and vertex selection       & 0.0--0.5     & 0.0--1.2    & --          & --         & -- \\
                \pT resolution                     & 0.0--3.0     & 0.0--1.4    & 0.0--3.0    & 0.0--2.7   & 0.0--1.0\\
                Interaction rate                   & --           & 0.0         & --          & --         & 1.0\\
                Material budget                    & 0.1--0.9     &0.1--0.9     & 0.1--0.9    & 0.1--0.9   & 0.1--0.9\\
                Acceptance                         & --           &--           & 0.0--0.2    & --         & --\\
                \hline
                Combined Uncertainty               &3.5--6.2      & 1.3--4.3    & 1.7--5.1    & 1.9--5.2   & 1.0--7.5\\
                \hline
                Normalization                      &1.9           & 2.3         & 3.1         & --         & -- \\
                Centrality                         & --           & --          & --          & 0.1--3.6   & 0.1--3.5\\
                \hline
                \end{tabular}
                \caption{Contributions to the relative systematic
                  uncertainty for \pt spectra in pp, p--Pb, and Pb--Pb
                  collisions. The ranges correspond to the maximal
                  variation within the considered \pt range of \mbox{0.15--50~GeV/$c$}, as well as Pb--Pb centrality intervals. The \pt-dependent contributions are assumed to be independent and are summed in quadrature, resulting in the combined uncertainty. All values are in \%.
                \label{tab:sys}}
\end{table}

\paragraph{Transverse momentum resolution}
The transverse momentum of charged particles is reconstructed from the
track curvature measured in the ITS and the TPC (see
\cite{Abelev:2014ffa} for details).  The modification of the spectra
arising from the finite momentum resolution is estimated from the
error obtained from the corresponding covariance matrix element of the Kalman fit. 
The relative \pt resolution, $\sigma(\pt)/\pt$,
depends on momentum and is approximately  3--4\% at $\pt = 0.15$~GeV/$c$, has
a minimum of 1.0\% at $\pt=1.0$~GeV/$c$, and increases linearly for
larger \pt, approaching 3--10\% at 50~GeV/$c$, depending on collision energy,
system or Pb--Pb centrality interval.  The \pt resolution has been
verified by studying the widths of the invariant mass distributions of $K^{{\rm 0}}_{{\rm s}}$ reconstructed from their decays to two charged
pions. 

To account for the finite \pT resolution, correction factors to the spectra were determined based on the Bayesian unfolding approach \cite{DAgostini:1994fjx} implemented in the RooUnfold package \cite{Adye:2011gm}. This unfolding is based on the response matrix, ${\bf R}^{\rm det}_{\rm m, t}$, which relates the measured spectrum ${\bf M}_{\rm m}$ and the true spectrum ${\bf T}_{\rm t}$,  \mbox{${\bf M}_{\rm m} = {\bf R}^{\rm det}_{\rm m, t} \cdot  {\bf T}_{\rm t}$}, where $m$ and $t$ are indices indicating the bin number. The response matrix was generated for each data set and Pb--Pb collision centrality using GEANT3 detector simulations with different Monte-Carlo generators. For $\pT > 10$~GeV/$c$, another unfolding procedure similar to what was done in previous work \cite{Abelev:2013ala} was also used. 

The correction factors depend on the collision energy and system as well as on the collision
centrality, due the change of the spectral shape.  For momenta below 10~GeV/$c$, the corrections are significant only in the first momentum bin $\pt = 0.15$--0.2~GeV/$c$, and reach 3\%(2.5\%) for pp(Pb--Pb) at $\snn = 2.76$~TeV,  3\% for p--Pb at $\snn = 5.02$~TeV and around 1\% for pp(Pb--Pb) at $\snn = 5.02$~TeV. At low \pt, these corrections are independent of Pb--Pb collision centrality. For $\pt>10$~GeV/$c$, both unfolding methods yield almost identical correction factors. For $\snn = 5.02$~TeV, the correction factors reach 5\%, 1.5\% and 3\% (4\%) at $\pt=50$~GeV/$c$ for pp, p--Pb and 0-5\%(70--80\%) central Pb--Pb collisions, respectively. For $\snn = 2.76$~TeV, they amount to 4\% for pp and 4\% (8\%) for 0-5\%(70--80\%) central Pb--Pb
collisions at the highest \pt. The resulting \pt-dependent correction factors are
applied (bin-by-bin) to the measured \pt spectra. 

\paragraph{Trigger and vertex selection}
The event selection (trigger and vertex) introduces a small $\pt$-dependence in the correction on 
the $\pt$ spectra in pp collisions. This is due to the fact that the low-multiplicity pp events, which are also characterized by a softer spectrum,
are mostly rejected by the trigger and vertex selection criteria.
The effect on the \pt spectra was calculated from simulations
with the PYTHIA 8 (Monash 2013 tune) and the PYTHIA 6 (Perugia2011 tune) event
generators and was estimated to be around 0.4--0.6\% (2.2--2.6\%) for $\pt < 1$~GeV/$c$
at \s = 2.76 (5.02) TeV. The spectra are corrected by the
average bias of these two generators, resulting in 0.5\% (2.4\%)
corrections to the spectra.

\paragraph{Acceptance correction for the p--Pb data}
The two-in-one magnet design of the LHC imposes the same magnetic
rigidity of the beams in the two rings. The configuration for p--Pb
collisions with protons at 4 TeV energy colliding with
$^{208}_{82}$Pb ions at 82$\times$4 TeV results in a shift in the
rapidity of the nucleon--nucleon center-of-mass system by $\Delta
y_{\text{NN}}$ = 0.465 in the direction of the proton beam (negative
$z$-direction). Therefore the detector coverage
$|\eta_{\text{lab}}|<0.8$ corresponds to roughly $-0.3
<\eta_{\text{cms}} < 1.3$.  For massless or high \pt particles,
$\eta_{\text{cms}} = \eta_{\text{lab}} + y_{\text{NN}}$ but the differential
yield of non-massless particles at low \pt suffers from a distortion,
which is estimated and corrected for based on the HIJING event
generator weighted by the measured relative particle abundances
\cite{Abelev:2013haa,Adam:2016dau}.  For \pt = 0.5 GeV/$c$ the
correction is 2\% for $-0.3 <\eta_{\text{cms}} < 1.3$. 

\subsection{Systematic uncertainty}
The relative systematic uncertainties on the $\pT$ spectra are
summarized in Table \ref{tab:sys}.
\begin{itemize} 
\item The effect of the selection of events based on the vertex position is
studied by comparing the fully corrected \pT spectra obtained with
alternative vertex selections corresponding to $\pm$5 and $\pm$20 cm.
\item The systematic uncertainties related to the track selection criteria
(listed above) were studied by varying the track quality cuts. In
particular, we varied the upper limits of the track fit quality
parameters in the ITS ($\chi_{{\rm ITS}}^{2}/{N_{\rm hits}}$) and the TPC ($\chi_{{\rm TPC}}^{2}/{N_{\rm clusters}}$)  
 in the ranges of 25--49 and
3--5, respectively. The systematic uncertainties related to high-\pt
fake tracks  \cite{Abelev:2012hxa}  were estimated by modifying the upper limits of the track
matching criteria given by the $\chi^{2}_{\rm TPC-ITS}$ in the range
of 25--49. The resulting uncertainty dominates at high \pt for all
collision systems.
\item The systematic uncertainty on the secondary-particle contamination
(Fig.~\ref{fig:eff}, right) includes contributions from the template
fits to the measured impact parameter distributions. We have varied
the fit model using templates with two (primaries, secondaries) or
three (primaries, secondaries from material, secondaries from weak
decays of $K^{0}_{{\rm s}}$ and $\Lambda$) components, as well as the
fit ranges. The maximum difference between the data and the 2
component-template fit is summed in quadrature with the difference
between results obtained from the 2 and the 3 component-template fits and result
is assigned as the systematic uncertainty on the contamination. This
contribution dominates for the lowest \pt independently of the collision
system.
\item The systematic uncertainty on the primary particle composition
consists of several contributions, including the extrapolation of the
spectra to low \pt, the approximation of the relative particle
abundances at high \pt, the efficiency parameterization at high \pt,
the uncertainties of the measured particle spectra and the MC
assumptions on the $\Sigma^{\pm} / \Lambda$ spectra ratios. For the
extrapolation to low \pt, we have studied different parameterizations
(Bylinkin and Rostovtsev, modified Hagedorn \cite{Hagedorn:1983tyu}, Blast-Wave) and
fit ranges. We have varied the \pt thresholds for the approximation of
the relative particle abundances as well as the efficiency
parameterization at high \pt. The measured particle spectra were
varied within systematic uncertainties (one particle species at a
time), and the resulting differences to the nominal spectra were
added in quadrature to the systematic uncertainties. We have also
assigned an additional uncertainty related to the different spectral
shape of $\Sigma^{\pm}$ and $\Lambda$ from the MC generators.
\item To account for the imperfect description of the experimental setup in simulations,
we compared the track matching between the TPC and the ITS
information in data and Monte Carlo after scaling of the fraction of
secondary particles obtained from the fits to the $d_{xy}$
distributions. After rescaling the fraction of secondary particles,
the agreement between data and Monte Carlo is within 4\%. This
value is assigned as an additional systematic uncertainty.

\item The systematic uncertainty on the \pt resolution at low \pt (only first \pt bin) was estimated by changing Monte-Carlo 
generators in the unfolding procedure. The pp collisions were simulated with PYTHIA and PHOJET,  p--Pb collisions with HIJING and DPMJET, and Pb--Pb collisions with HIJING and AMPT \cite{Lin:2004en}. The average correction factor of two generators was assigned as systematic uncertainty. At low \pt, we observe a weak dependence of correction factors on the considered Monte-Carlo generators. The resulting uncertainties amount to 3\%(2.5\%) for pp(Pb--Pb) collisions at $\snn = 2.76$~TeV, to 3\% for p--Pb collisions at $\snn = 5.02$~TeV, and to 1\% for pp and Pb--Pb collisions at $\snn = 5.02$~TeV. 
The systematic uncertainty on the \pt resolution at high \pt ($>10$~GeV/$c$) was estimated using
the azimuthal angle dependence of the $1/\pt$ spectra for positively
and negatively charged particles. The relative shift of the spectra
for oppositely charged particles along $1/\pt$ determines the size of
uncertainty for a given angle. We used the RMS of the 1/\pt shift
distribution for the full azimuth as additional smearing of the \pt
resolution. We checked that these shifts are due to detector effects (such as ${\bf E} \times {\bf B}$ effect) and 
are not related to the physics of hadronic interaction in GEANT3.
To take into account the decrease in the \pt resolution with
increasing interaction rate, we have studied the systematic
uncertainty for the pp and Pb--Pb data sets at \snn = 5.02~TeV,
obtained from the difference of the spectra at high and low
interaction rate. The uncertainty is negligible for pp collisions, and is about 1\% for Pb--Pb collisions.
\item For the correction due to the trigger and vertex selection, calculated as
the average bias of two generators, half of the value is assigned as
systematic uncertainty.
\item The systematic uncertainty for the acceptance correction on the p--Pb
data was estimated by varying the relative particle abundances within
their measured uncertainties and by changing the fit function for the
low-\pt extrapolation. The uncertainty is sizable only at low \pt
where it reaches 0.2\%.
\item The material budget in the simulation was varied by
$\pm$4.5\% \cite{Abelev:2014ffa}, resulting in the systematic uncertainty in the range of 0.1--0.9\%.
\item The   normalization uncertainty on the
spectra in pp collisions was propagated from the cross section
measurements.  
\item The systematic uncertainties related to centrality
selection were estimated by a comparison of the \pT spectra when the
limits of the centrality classes are shifted due to an uncertainty of
$\pm$ 0.5\% in the fraction of the hadronic cross section used in the
analysis and by a comparison of results obtained using the SPD detector 
to estimate centrality as opposed to the V0A and V0C. 
\end{itemize} 

For the evaluation of the total systematic uncertainty all
contributions are considered to be uncorrelated and
 they are summed in quadrature. The improved
reconstruction and track selection in the reanalysis of pp and Pb--Pb
data at $\snn = 2.76$~TeV and p--Pb data at $\snn = 5.02$~TeV lead to
significantly reduced systematic uncertainties by a factor of about 2 as compared to
previously published results \cite{Abelev:2012hxa,Abelev:2013ala,Abelev:2014dsa}.

\section{Results and discussion}
\subsection{Spectra}
\begin{figure}[htb]
  {\centering 
\resizebox*{.49\columnwidth}{!}{\includegraphics{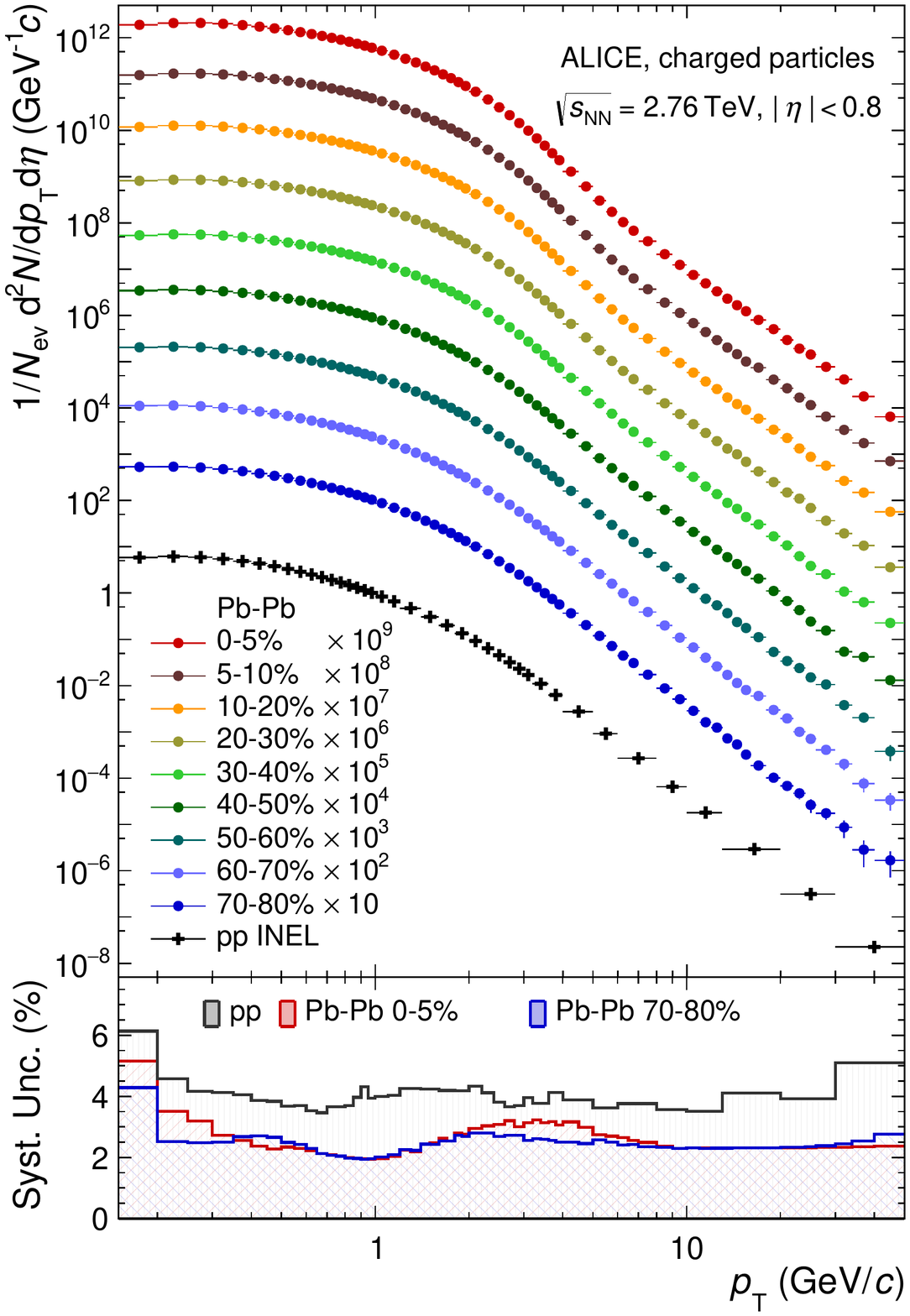}} 
\resizebox*{.49\columnwidth}{!}{\includegraphics{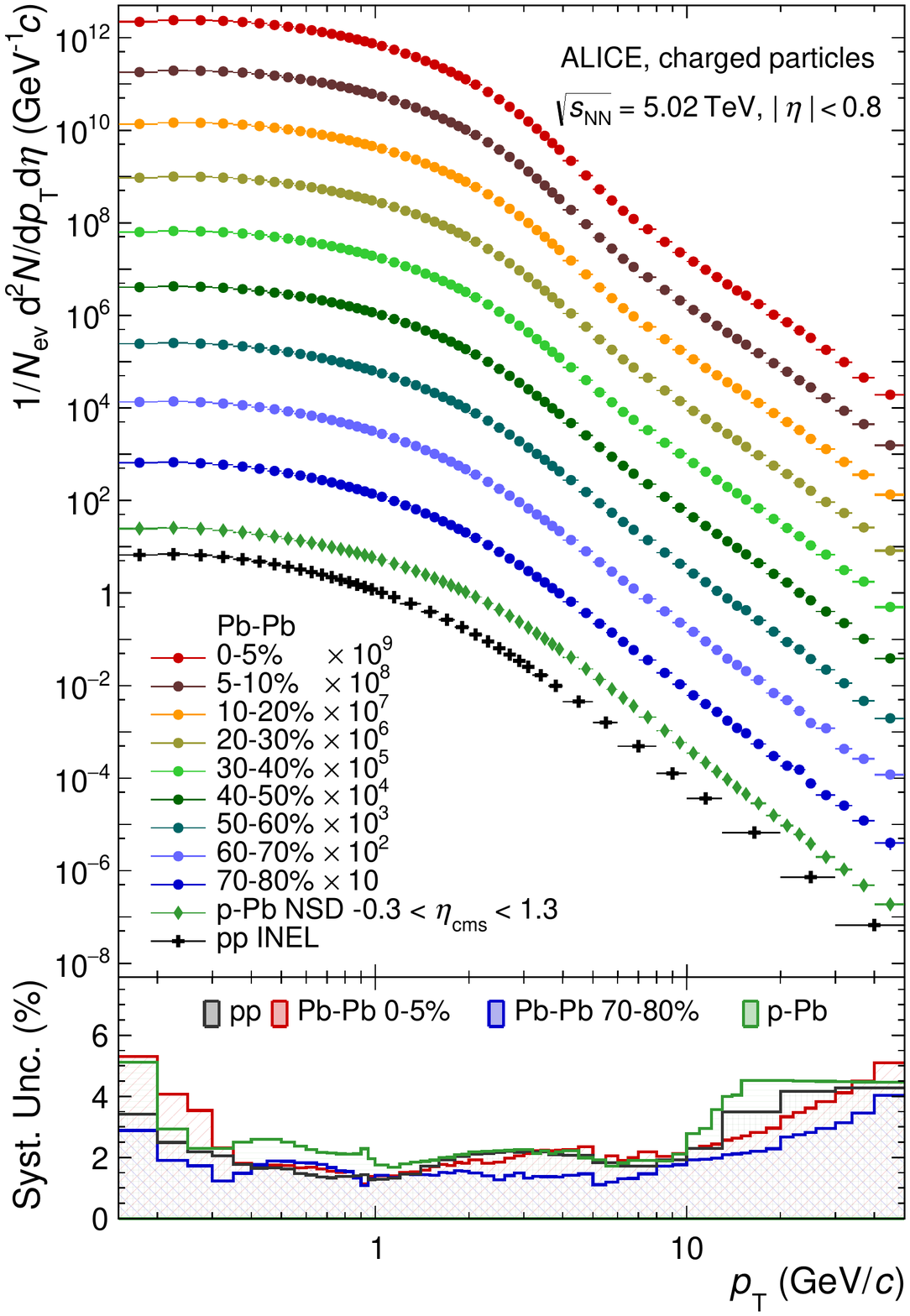}} 
\par}
\caption{\label{fig:spectra} Transverse momentum distributions of
  primary charged particles in $|\eta|<0.8$ in nine centrality
  intervals in Pb--Pb collisions at \snn = 2.76 (left) and 5.02 TeV
  (right) (scale factors as indicated are used for better
  visibility). The data for pp collisions, obtained scaling the cross
  section by $\sigma^{\rm NN}_{\rm inel}$, and NSD p--Pb at
  $\snn~=~5.02$~TeV are also shown.  The relative systematic
  uncertainties are shown in the lower panels for various datasets;
  these do not contain the normalization uncertainty.}
\end{figure}

The fully corrected \pt spectra of primary charged particles measured
in INEL  pp and Pb--Pb collisions at $\snn = 2.76$~TeV and 5.02~TeV and in NSD
p--Pb collisions at $\snn = 5.02$ are shown in
Fig.~\ref{fig:spectra}. The \mbox{Pb--Pb} spectra are presented in nine
centrality classes.
For pp collisions, the $\pt$-differential cross sections
are divided by the corresponding inelastic nucleon-nucleon cross
section at $\s=2.76$ (61.8~mb) and 5.02~TeV (67.6~mb) \cite{Loizides:2017aya}, respectively.
The relative systematic uncertainties for the various datasets are
shown in the bottom panels.  Substantial improvements in track
selection and efficiency corrections have been achieved. However the
uncertainty on the pp data at $\snn = 2.76$~TeV is still larger
than for the data at \snn = 5.02~TeV due to larger number of inactive channels in the SPD \cite{Abelev:2014ffa}, 
which affects the track reconstruction and the determination of the secondary particle contribution.

In Pb--Pb collisions the shape of the \pt spectrum varies strongly
with collision centrality. For peripheral collisions, the spectral shape is similar to
that measured in pp collisions as well as to the spectrum in p--Pb
collisions. With increasing collision centrality, a marked depletion
of the Pb--Pb spectra develops for $\pt > 5$~GeV/$c$. 
 These measurements supersede our previous
results \cite{Abelev:2012hxa,Abelev:2013ala,Abelev:2014dsa}, which allows for a better discrimination
between jet quenching scenarios.

\begin{figure}[htb]
\centering 
\resizebox*{.7\columnwidth}{!}{\includegraphics{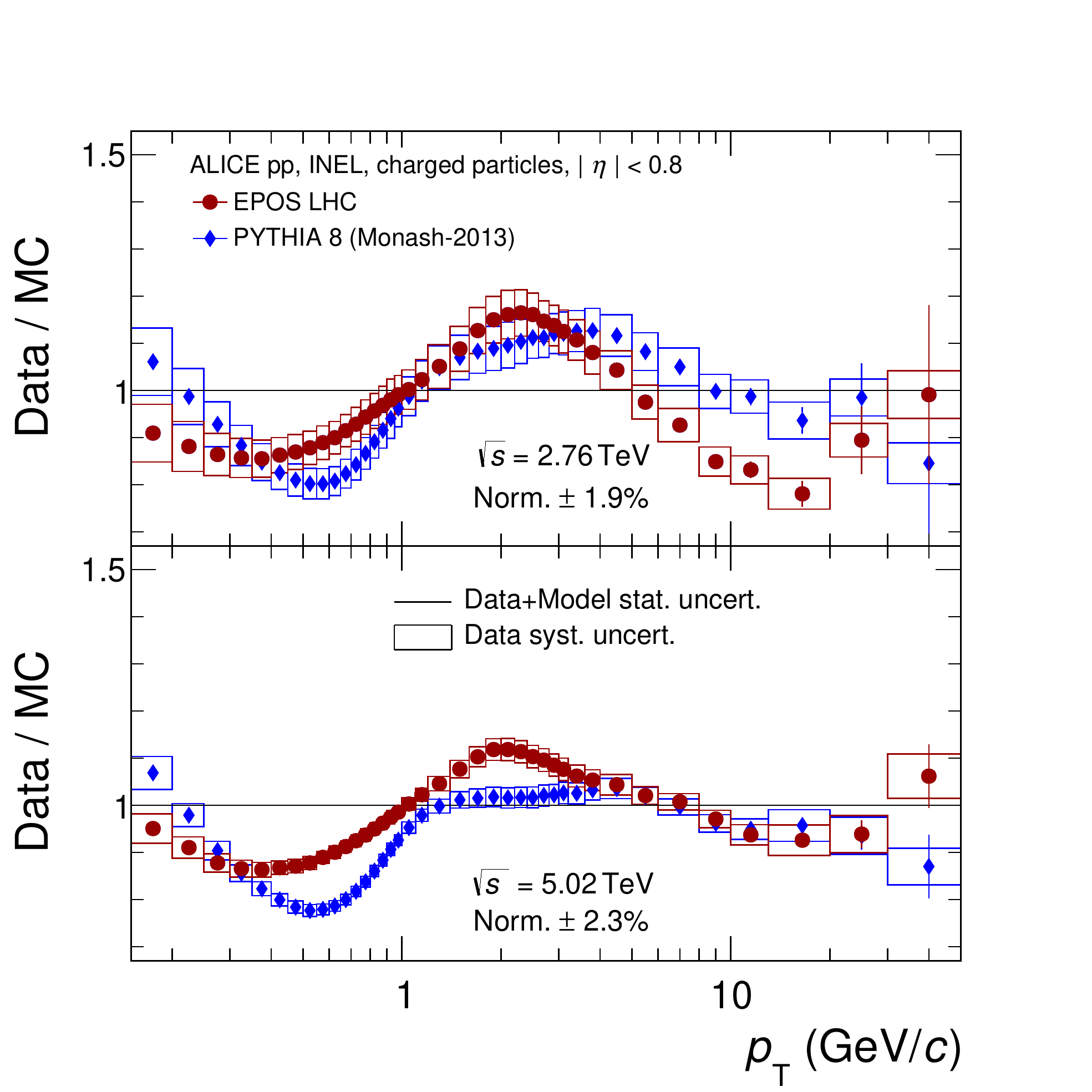} \par}
\caption{\label{fig:ppmodel} Comparison of the charged-particle transverse momentum spectra
 measured in pp collisions to PYTHIA 8  (Monash-2013 tune) \cite{Sjostrand:2008aro, Skands:2014mon}  and EPOS \cite{Pierog:2013ria} model
  calculations at \s = 2.76 (top) and 5.02 TeV (bottom). The statistical uncertainties of the data and model calculations are added in quadrature. The boxes represent systematic uncertainties of the data.}
\end{figure}

Figure \ref{fig:ppmodel} compares the measured \pt spectra in pp
collisions with results from PYTHIA 8 (Monash-2013 tune),
including colour reconnection, and EPOS LHC \cite{Pierog:2013ria}, which incorporates
collective (flow-like) effects. These event generators show a similar
description of the \pt spectra at both energies. They reproduce the spectral shape within 20\%.  

\begin{figure}[htb]
  {\centering 
\resizebox*{.7\columnwidth}{!}{\includegraphics{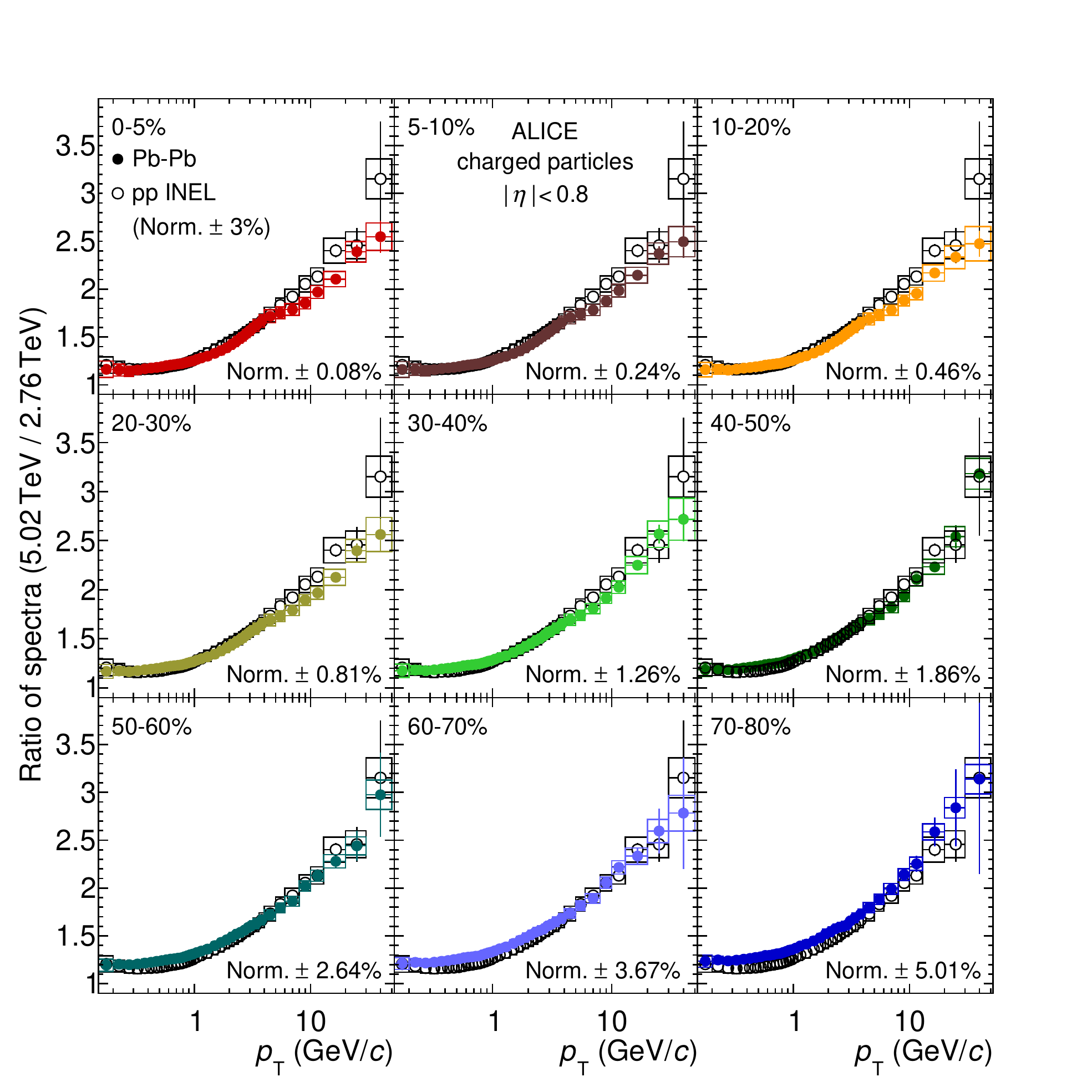}} 
\par}
\caption{\label{fig:ratio} Ratio of transverse momentum spectra at
  \snn = 5.02 and \snn = 2.76 TeV for Pb--Pb collisions, for nine
  centrality classes, and in pp collisions (repeated in each
  panel). The relative normalization uncertainties due to the
  centrality determination are indicated for each centrality class. For the pp spectrum,  the relative normalization uncertainty is $\pm 3\%$.}
\end{figure}

Figure \ref{fig:ratio} shows the ratios of \pt spectra measured at
\snn = 5.02 and \snn = 2.76 TeV in Pb--Pb and pp collisions. The
ratios for Pb--Pb collisions are determined in nine centrality classes ranging
from 0--5\% (top-left) to the 70--80\% (bottom-right). As
indicated by the ratios, the \pt spectra measured at higher collision
energy are significantly harder for both Pb--Pb and pp collision
systems. One can see that there is a similar energy
dependence of the ratio for peripheral (70--80\%) Pb--Pb and in pp
collisions, while towards central Pb--Pb collisions a gradual
reduction of the ratio is apparent. 

\begin{figure}[htb]
  {\centering 
\resizebox*{.7\columnwidth}{!}{\includegraphics{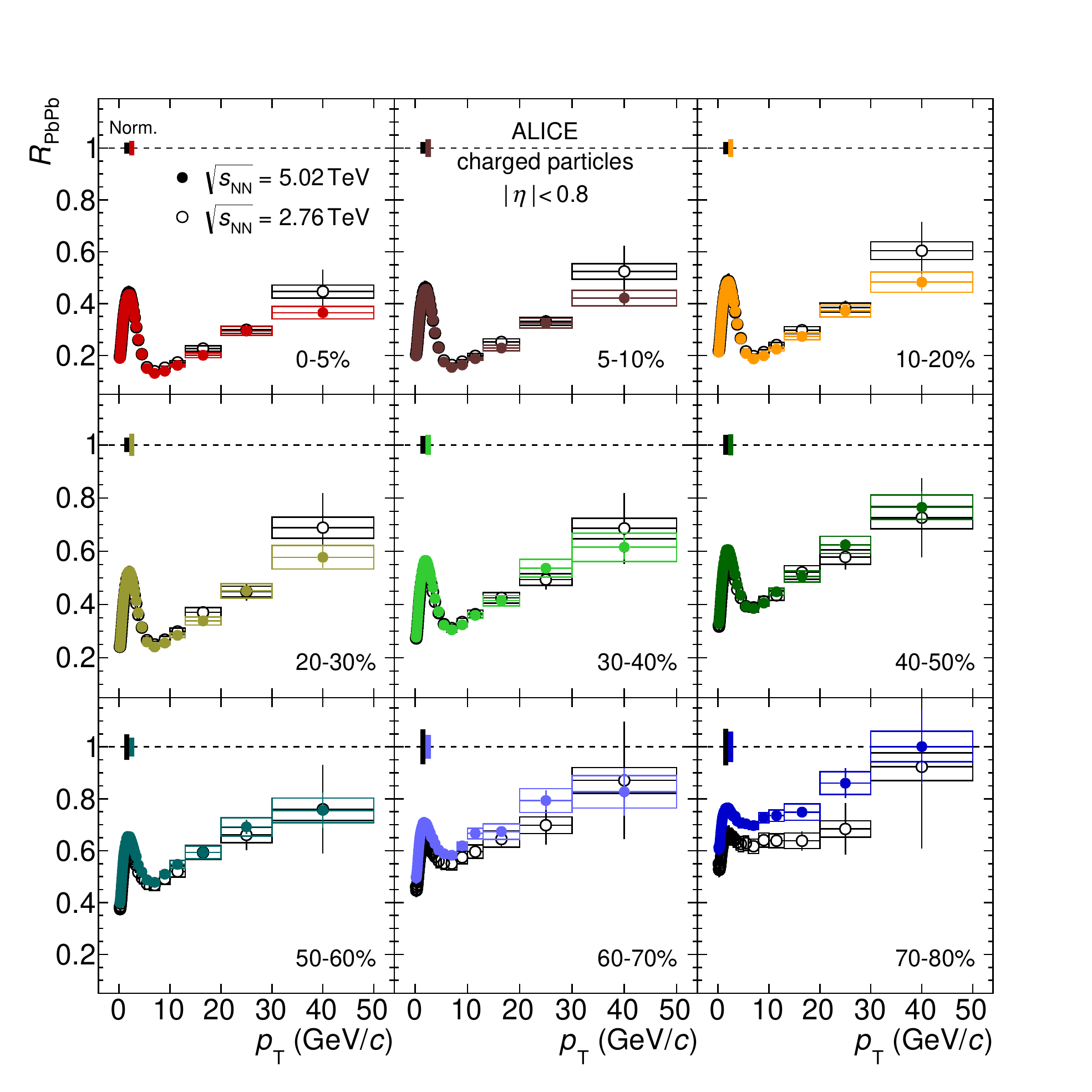}} 
\par}
\caption{\label{fig:raa} The transverse momentum dependence of the
  nuclear modification factor measured in Pb--Pb collisions, for nine centrality classes.  The
  new data at \snn = 5.02 (full symbols) are compared to the
  reanalyzed data at \snn = 2.76 TeV (open symbols). The normalization uncertainties are shown as boxes around unity.}
\end{figure}

\subsection{Nuclear modification factors}
In order to quantify in-medium modification of charged-particle transverse momentum spectrum, 
the nuclear modification factors are determined.  Figure~\ref{fig:raa} shows the \raa for Pb--Pb
collisions measured at \snn = 2.76 and 5.02
TeV. The nuclear modification factor has a strong centrality
dependence, and is very similar in magnitude for the two collision
energies. 

Given that the \pt spectra are harder at the higher \snn (see Fig.~\ref{fig:ratio}) and that the medium density increases with \snn by $\sim20\%$ \cite{Adam:2015ptt}, 
this similarity of the \raa may indicate a larger parton energy loss in the hotter/denser and longer-lived deconfined medium produced at the higher center-of-mass energy.  Assuming that the initial parton \pt spectrum, parton distribution and fragmentation functions are not significantly modified by the energy increase, and that the parton energy loss in expanding medium is sublinear to the medium density increase, we would expect larger energy loss at $\snn = 5.02$~TeV than at  $\snn = 2.76$~TeV, but no more than 20\%.

In  0--5\% central collisions the yield is suppressed by a
factor of about 8 ($\raa \approx 0.13$) at
\mbox{$\pt=6$--7~GeV/$c$}. Above $\pt=7$~GeV/$c$, there is a
significant rise of the nuclear modification factor, which reaches a
value of about 0.4 for our highest \pt bin, 30--50~GeV/$c$. In
peripheral collisions (70--80\%), the suppression is 30\% for
intermediate momenta and approaches unity for the highest \pt bin.
The normalization uncertainties for \raa originate from the pp
measurement and centrality determination and were added in
quadrature.

\begin{figure}[htb]
  {\centering 
\resizebox*{.49\columnwidth}{!}{\includegraphics{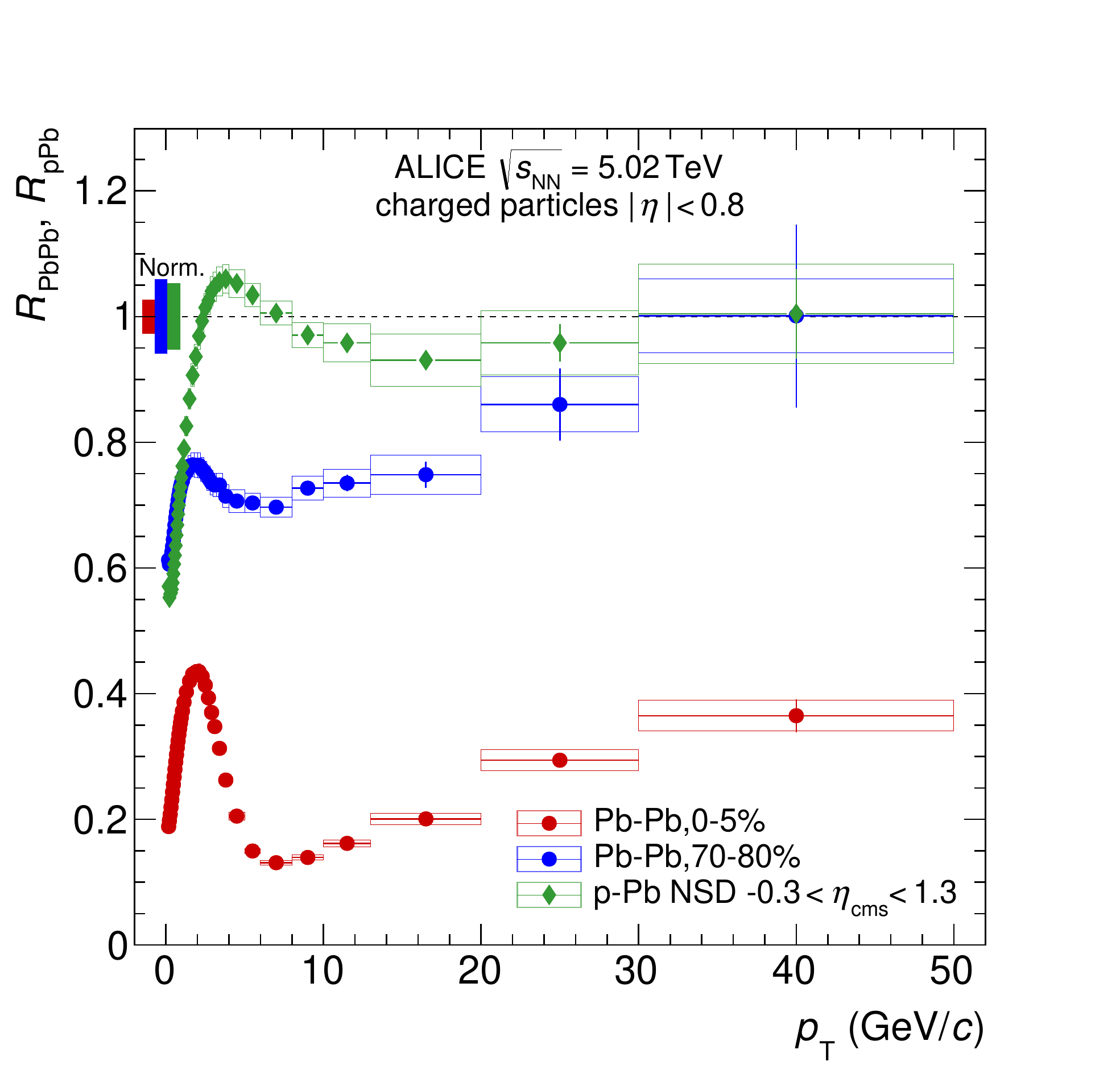}} 
\resizebox*{.49\columnwidth}{!}{\includegraphics{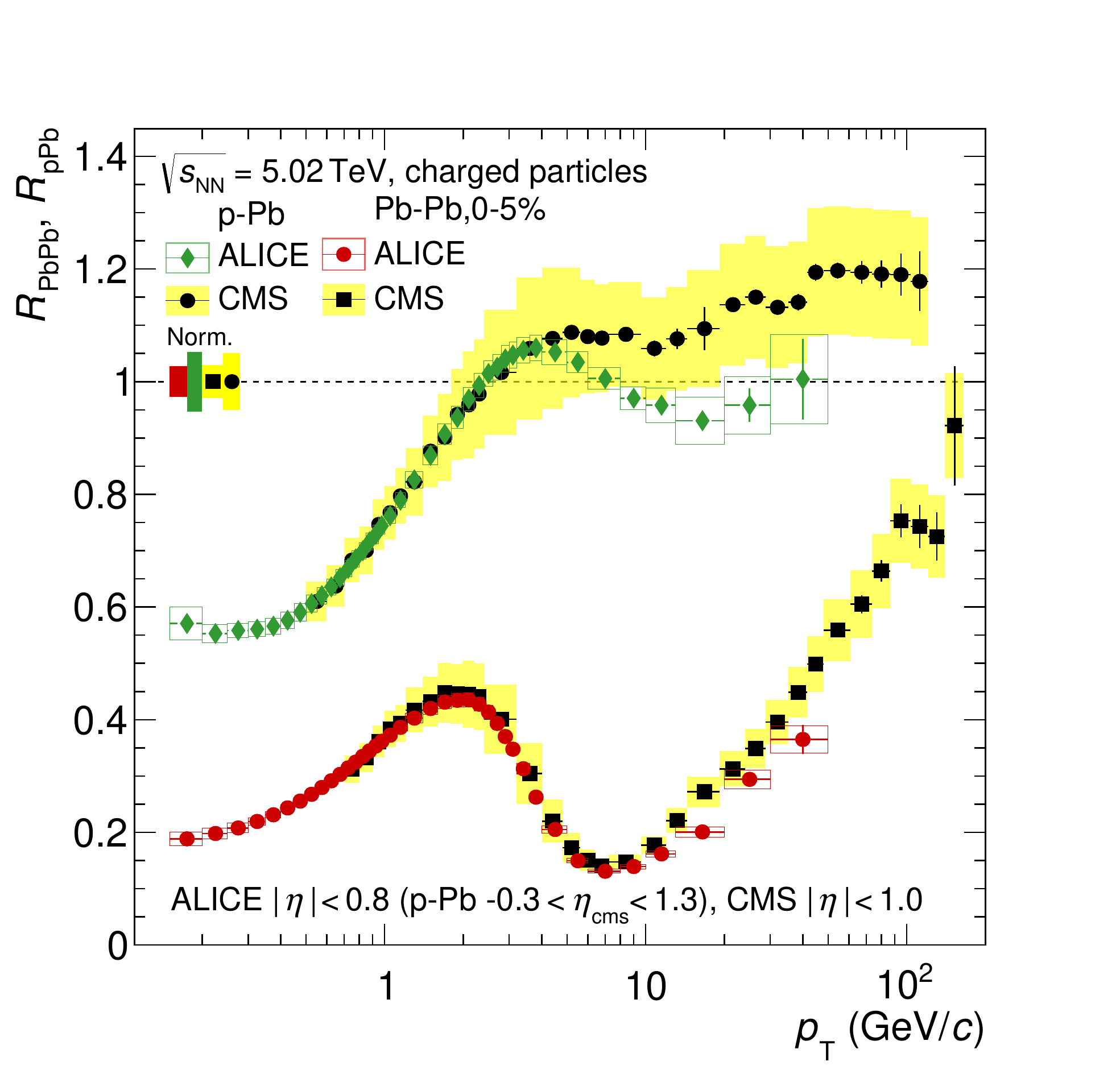}} 
\par}
\caption{\label{fig:raa_rpa} Left: Nuclear modification
  factors measured by ALICE in central (0--5\%) and peripheral
  (70--80\%) Pb--Pb collisions and in p--Pb collisions at $\snn =
  5.02$ TeV.  Right: A comparison of the nuclear modification
  factors for central (0-5\%) Pb--Pb and p--Pb collisions measured by
  ALICE and CMS \cite{Khachatryan:2015xaa,Khachatryan:2016odn}. 
  In both figures, the \pt-dependent systematic uncertainties are shown as
  boxes around data points. The normalization uncertainties are shown
  as boxes around unity.}
\end{figure}

Figure \ref{fig:raa_rpa} (left) shows the \rppb factor compared
to \raa measured in the 0--5\% and 70--80\% centrality classes for
Pb--Pb collisions at $\snn = 5.02$. The \rppb factor exhibits a maximum 
for the intermediate \pt range, $2<\pt<6$~GeV/$c$, a
feature generically called the Cronin effect
\cite{Cronin:1975atg}. 
A study on its dependence on the particle species
\cite{Abelev:2013haa} suggested that protons are responsible for the
observed maximum. The maximum occurs at values of
\pt (3--5 GeV/$c$) larger than the maximum of \raa seen in the \pt
range 1.5--3 GeV/$c$.
The \rppb factor is consistent with unity for $\pt\gtrsim8$ GeV/$c$,
demonstrating that the strong suppression observed in central Pb--Pb
collisions is not related to initial state effects but rather to the
formation of hot and dense QCD matter.  The ALICE results for \raa and
\rppb measured at $\snn = 5.02$~TeV are compared to measurements by
CMS \cite{Khachatryan:2015xaa} in Fig. \ref{fig:raa_rpa} (right).  Agreement within $1.5\sigma$
is observed for both \raa and \rppb taking into account the current uncertainties.

\begin{figure}[htb]
  {\centering 
\resizebox*{.7\columnwidth}{!}{\includegraphics{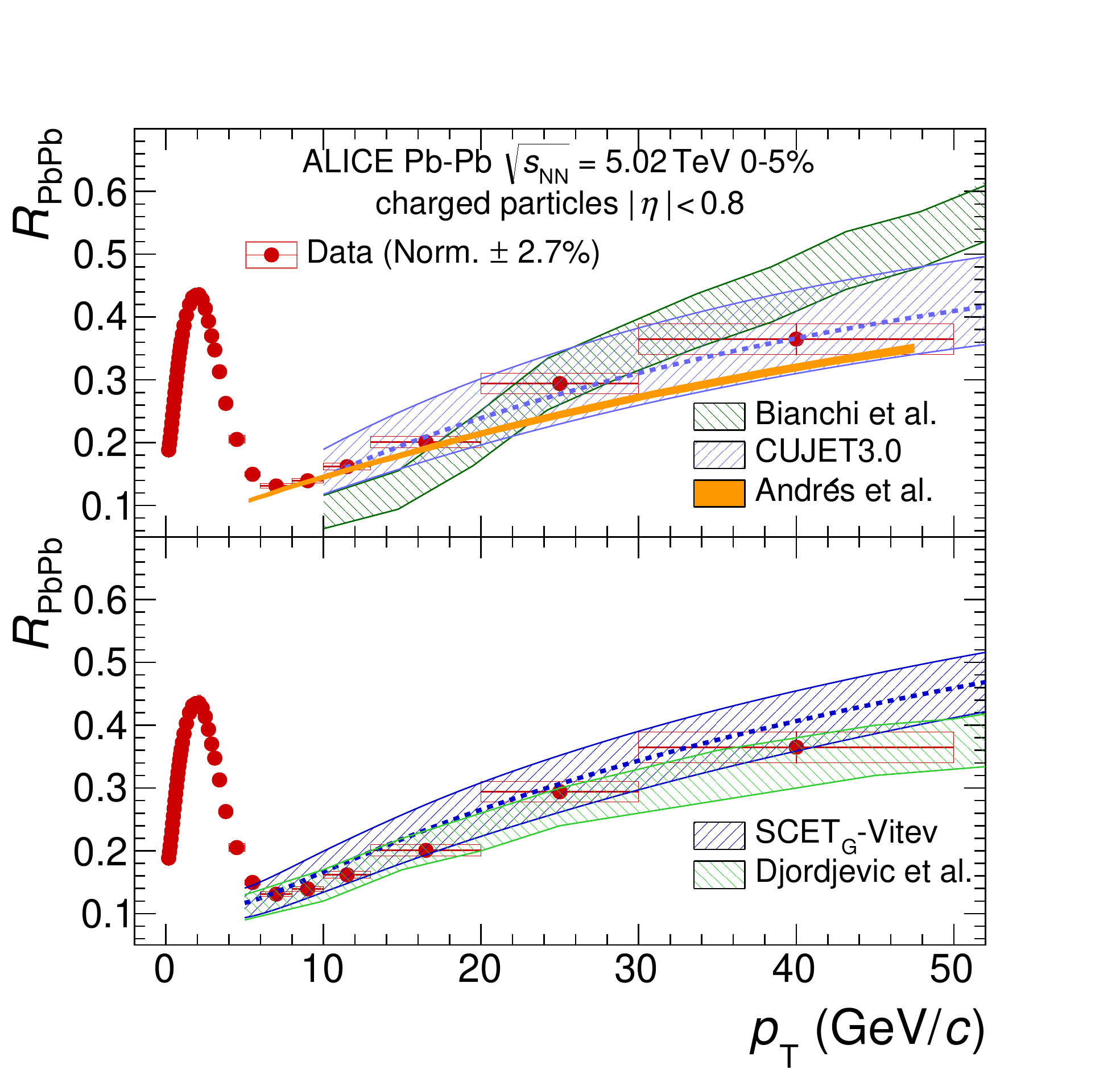}} 

\par}
\caption{\label{fig:raa_models} The charged-particle nuclear
  modification factor measured in the 0--5\% most central
  Pb--Pb collisions at $\snn = 5.02$~TeV in comparison to model predictions
  \cite{Kang:2014xsa, Chien:2016adr, Djordjevic:2015hra, Djordjevic:2016vfo} (lower
  panel) and \cite{Bianchi:2017rit, Xu:2015arg, Xu:2016eru,
    Andres:2016art} (upper panel). The red boxes around data points
  represent \pt dependent systematic uncertainties. The normalization
  uncertainty of the data ($\pm 2.7\%$) is not part of the
  uncertainties of the plotted data points.
} 
\end{figure}

\subsection{Comparison with theoretical models}
In Fig.~\ref{fig:raa_models} the measured \raa for 0-5\% 
 central collisions at \snn = 5.02~TeV is compared to model predictions. All presented models are based on the pQCD factorization, where the entire effect of energy loss is encoded in the medium-modified parton fragmentation function.  All models include radiative energy loss based on different approaches. The model by Djordjevic et al. \cite{Djordjevic:2015hra, Djordjevic:2016vfo}  and CUJET~3.0 \cite{Xu:2015arg, Xu:2016eru} include in addition collisional energy loss. The energy loss is calculated in dynamically expanding medium in all models except that of Vitev et al. \cite{Kang:2014xsa, Chien:2016adr}, in which the medium is composed of static scattering centers. In the following, the models are discussed in more detail. 
 
The calculations by Vitev et al. are based on the SCET$_{\rm G}$ model
\cite{Kang:2014xsa, Chien:2016adr}, which uses an extended soft-collinear effective
theory to describe inclusive particle production and suppression in
the heavy-ion environment. This theoretical framework provides an
analytic connection between generalized DGLAP evolution equations for
the fragmentation functions in dense strongly-interacting matter and
parton energy loss for hard processes.  The calculations employ the
pQCD-based hard cross section and QGP medium evolved parton-to-hadron
fragmentation functions, combined with initial-state cold nuclear
matter (CNM) effects, which include dynamical nuclear shadowing, the
Cronin effect and initial-state parton energy loss (see 
\cite{Chien:2016adr} and references therein for details). The two upper and lower
curves represent calculations for the nuclear modification factor with
variations of the coupling strength $g = 1.9 \pm 0.1$ between the jet and the medium, which is a free parameter in the calculations.
Djordjevic et al. \cite{Djordjevic:2015hra, Djordjevic:2016vfo} use a
dynamical energy loss formalism based on pQCD calculations in a
finite size dynamical QCD medium. While the initial \pt spectrum is
the same as that used in the SCET$_{\rm G}$ model, the dynamical
description of the medium provides a consistent treatment of both
radiative and collisional energy loss, including a finite magnetic
screening mass, which modifies the gluon self energy and therefore
changes the energy loss, as well as a running coupling constant for
the strongly-interacting medium. The two curves correspond to
different electric-to-magnetic screening mass ratios in the range
$0.4<\mu_{{\rm M}}/\mu_{{\rm E}}<0.6$.
The model of Bianchi et al. \cite{Bianchi:2017rit} uses the pQCD
factorization scheme with a pQCD-based radiative energy loss
in a hydrodynamically expanding medium. In this
framework, high \pt hadrons arise from fragmentation of hard
partons, which lose energy prior to hadronization via interactions with
the medium. The amount of energy loss is regulated by the medium
transport coefficient $\hat{q}$, which varies with the
temperature-dependent entropy density of the medium as well as with
the energy scale of jets propagating in the medium.
The CUJET 3.0 model \cite{Xu:2015arg, Xu:2016eru} is an extension of the perturbative-QCD-based CUJET 2.0 model, with the two complementary non-perturbative features of the QCD cross-over phase transition: the suppression of quark and gluon degrees of freedom and the emergence of chromomagnetic monopoles. The calculations were performed varying the value of the QCD running coupling $\alpha_{{\rm c}}$ from 0.95 to 1.33 for $Q < T_{{\rm C}}$, and the ratio of electric to magnetic screening scales $c_{{\rm m}} = g_{\rm s} \mu_{{\rm E}} / \mu_{M}$ ($c_{{\rm m}} = 0,0.3, 0.4$), where $g_{\rm s}$ is the strong coupling constant. The value of $\alpha_{\rm c}$ was fixed for each $c_{{\rm m}}$ value by fitting a single reference datum, $\raa (\pt = 12~{\rm GeV}/c) \approx 0.3$, for charged hadrons in 20--30\% central Pb--Pb collisions at the LHC. 
The calculations by Andr{\'e}s  et al. \cite{Andres:2016art} use the jet quenching formalism of quenching weights. This approach consists of fitting a $K$ factor, defined as $ K \equiv \hat{q} / 2 \epsilon^{3/4}$, that quantifies departure of this parameter from the perturbative estimate, $\hat{q}_{\rm ideal} \sim 2 \epsilon^{3/4}$~\cite{Baier:2003rty}, where the local energy density $\epsilon$ is taken from a hydrodynamical model of the medium. The $K$ factor is the only free parameter in the fit of nuclear modification factors. Without including new data at $\snn = 5.02$~TeV in the fit procedure, they predict a $\sim15\%$ larger suppression at  $\snn = 5.02$~TeV as compared to $\snn = 2.76$~TeV, assuming the same value of K as the one obtained from the fit to the data at the lower energy.

All models presented here describe the main features of the data. The models by Vitev et al., Djordjevic et al. and CUJET 3.0 give quantitatively good description of the data. The model by Bianchi et al. is consistent with data within $1.5\sigma$ while that by Andr{\'e}s et al. underestimates the data at high $\pt$. However, one should note that this comparison is made between unbinned theory calculations and binned data in relatively large \pt bins, which might introduce additional uncertainty. 

\section{Summary}

In summary, we measured the primary charged particle \pT\ spectra in pp and Pb--Pb collisions at $\snn = 5.02$~TeV. We also reanalyzed the data collected in pp and Pb--Pb collisions at $\snn=2.76$~TeV as well as in p--Pb collisions at $\snn = 5.02$~TeV with the revised techniques. Thanks to an improved reconstruction, track selection and data-driven efficiency correction procedure we were able to reduce the systematic uncertainties by a factor of $\sim 2$ as compared to previously published ALICE results. The measured spectra were used to determine the nuclear modification factors $\rppb$ and $\raa$.
The nuclear modification factor in p--Pb collisions is consistent with unity at high \pT,
showing that the strong suppression observed in Pb--Pb is not due
to CNM effects but rather due to final state partonic energy loss in
the hot and dense QGP created in Pb--Pb collisions.  This
suppression is weak in peripheral collisions
and increases with centrality reaching a value of $\raa=0.13$ at
\pT = 6--7 GeV, indicating an increasing parton energy loss with
centrality. This suppression is found to be similar at $\snn = 2.76$
and 5.02 TeV, despite the much harder \pT\ spectrum at the top energy, which may indicate
 a stronger parton energy loss and a larger energy density of
the medium at the higher energy. All models presented here describe the main features of the data with Vitev et al., Djordjevic et al. and CUJET 3.0 being compatible with data within uncertainties. However, further precision in the theoretical calculations is needed to extract the transport properties of the hot and dense deconfined QCD matter.

%
%

\newenvironment{acknowledgement}{\relax}{\relax}
\begin{acknowledgement}
\section*{Acknowledgements}

The ALICE Collaboration would like to thank all its engineers and technicians for their invaluable contributions to the construction of the experiment and the CERN accelerator teams for the outstanding performance of the LHC complex.
The ALICE Collaboration gratefully acknowledges the resources and support provided by all Grid centres and the Worldwide LHC Computing Grid (WLCG) collaboration.
The ALICE Collaboration acknowledges the following funding agencies for their support in building and running the ALICE detector:
A. I. Alikhanyan National Science Laboratory (Yerevan Physics Institute) Foundation (ANSL), State Committee of Science and World Federation of Scientists (WFS), Armenia;
Austrian Academy of Sciences and Nationalstiftung f\"{u}r Forschung, Technologie und Entwicklung, Austria;
Ministry of Communications and High Technologies, National Nuclear Research Center, Azerbaijan;
Conselho Nacional de Desenvolvimento Cient\'{\i}fico e Tecnol\'{o}gico (CNPq), Universidade Federal do Rio Grande do Sul (UFRGS), Financiadora de Estudos e Projetos (Finep) and Funda\c{c}\~{a}o de Amparo \`{a} Pesquisa do Estado de S\~{a}o Paulo (FAPESP), Brazil;
Ministry of Science \& Technology of China (MSTC), National Natural Science Foundation of China (NSFC) and Ministry of Education of China (MOEC) , China;
Ministry of Science, Education and Sport and Croatian Science Foundation, Croatia;
Ministry of Education, Youth and Sports of the Czech Republic, Czech Republic;
The Danish Council for Independent Research | Natural Sciences, the Carlsberg Foundation and Danish National Research Foundation (DNRF), Denmark;
Helsinki Institute of Physics (HIP), Finland;
Commissariat \`{a} l'Energie Atomique (CEA) and Institut National de Physique Nucl\'{e}aire et de Physique des Particules (IN2P3) and Centre National de la Recherche Scientifique (CNRS), France;
Bundesministerium f\"{u}r Bildung, Wissenschaft, Forschung und Technologie (BMBF) and GSI Helmholtzzentrum f\"{u}r Schwerionenforschung GmbH, Germany;
General Secretariat for Research and Technology, Ministry of Education, Research and Religions, Greece;
National Research, Development and Innovation Office, Hungary;
Department of Atomic Energy Government of India (DAE), Department of Science and Technology, Government of India (DST), University Grants Commission, Government of India (UGC) and Council of Scientific and Industrial Research (CSIR), India;
Indonesian Institute of Science, Indonesia;
Centro Fermi - Museo Storico della Fisica e Centro Studi e Ricerche Enrico Fermi and Istituto Nazionale di Fisica Nucleare (INFN), Italy;
Institute for Innovative Science and Technology , Nagasaki Institute of Applied Science (IIST), Japan Society for the Promotion of Science (JSPS) KAKENHI and Japanese Ministry of Education, Culture, Sports, Science and Technology (MEXT), Japan;
Consejo Nacional de Ciencia (CONACYT) y Tecnolog\'{i}a, through Fondo de Cooperaci\'{o}n Internacional en Ciencia y Tecnolog\'{i}a (FONCICYT) and Direcci\'{o}n General de Asuntos del Personal Academico (DGAPA), Mexico;
Nederlandse Organisatie voor Wetenschappelijk Onderzoek (NWO), Netherlands;
The Research Council of Norway, Norway;
Commission on Science and Technology for Sustainable Development in the South (COMSATS), Pakistan;
Pontificia Universidad Cat\'{o}lica del Per\'{u}, Peru;
Ministry of Science and Higher Education and National Science Centre, Poland;
Korea Institute of Science and Technology Information and National Research Foundation of Korea (NRF), Republic of Korea;
Ministry of Education and Scientific Research, Institute of Atomic Physics and Romanian National Agency for Science, Technology and Innovation, Romania;
Joint Institute for Nuclear Research (JINR), Ministry of Education and Science of the Russian Federation and National Research Centre Kurchatov Institute, Russia;
Ministry of Education, Science, Research and Sport of the Slovak Republic, Slovakia;
National Research Foundation of South Africa, South Africa;
Centro de Aplicaciones Tecnol\'{o}gicas y Desarrollo Nuclear (CEADEN), Cubaenerg\'{\i}a, Cuba and Centro de Investigaciones Energ\'{e}ticas, Medioambientales y Tecnol\'{o}gicas (CIEMAT), Spain;
Swedish Research Council (VR) and Knut \& Alice Wallenberg Foundation (KAW), Sweden;
European Organization for Nuclear Research, Switzerland;
National Science and Technology Development Agency (NSDTA), Suranaree University of Technology (SUT) and Office of the Higher Education Commission under NRU project of Thailand, Thailand;
Turkish Atomic Energy Agency (TAEK), Turkey;
National Academy of  Sciences of Ukraine, Ukraine;
Science and Technology Facilities Council (STFC), United Kingdom;
National Science Foundation of the United States of America (NSF) and United States Department of Energy, Office of Nuclear Physics (DOE NP), United States of America.    
\end{acknowledgement}

\bibliographystyle{utphys}   
\bibliography{raa}

\newpage
\appendix
\section{The ALICE Collaboration}
\label{app:collab}

\begingroup
\small
\begin{flushleft}
S.~Acharya\Irefn{org139}\And 
F.T.-.~Acosta\Irefn{org22}\And 
D.~Adamov\'{a}\Irefn{org94}\And 
J.~Adolfsson\Irefn{org81}\And 
M.M.~Aggarwal\Irefn{org98}\And 
G.~Aglieri Rinella\Irefn{org36}\And 
M.~Agnello\Irefn{org33}\And 
N.~Agrawal\Irefn{org48}\And 
Z.~Ahammed\Irefn{org139}\And 
S.U.~Ahn\Irefn{org77}\And 
S.~Aiola\Irefn{org144}
% ...
\And
\renewcommand\labelenumi{\textsuperscript{\theenumi}~}

\section*{Affiliation notes}
\renewcommand\theenumi{\roman{enumi}}
\begin{Authlist}
\item \Adef{org*}Deceased
\item \Adef{orgI}Dipartimento DET del Politecnico di Torino, Turin, Italy
\item \Adef{orgII}M.V. Lomonosov Moscow State University, D.V. Skobeltsyn Institute of Nuclear, Physics, Moscow, Russia
\item \Adef{orgIII}Department of Applied Physics, Aligarh Muslim University, Aligarh, India
\item \Adef{orgIV}Institute of Theoretical Physics, University of Wroclaw, Poland
\end{Authlist}

\section*{Collaboration Institutes}
\renewcommand\theenumi{\arabic{enumi}~}
\begin{Authlist}
\item \Idef{org1}A.I. Alikhanyan National Science Laboratory (Yerevan Physics Institute) Foundation, Yerevan, Armenia
\item \Idef{org2}Benem\'{e}rita Universidad Aut\'{o}noma de Puebla, Puebla, Mexico
\item \Idef{org3}Bogolyubov Institute for Theoretical Physics, National Academy of Sciences of Ukraine, Kiev, Ukraine
\item \Idef{org4}Bose Institute, Department of Physics  and Centre for Astroparticle Physics and Space Science (CAPSS), Kolkata, India
\item \Idef{org5}Budker Institute for Nuclear Physics, Novosibirsk, Russia
\item \Idef{org6}California Polytechnic State University, San Luis Obispo, California, United States
\item \Idef{org7}Central China Normal University, Wuhan, China
\item \Idef{org8}Centre de Calcul de l'IN2P3, Villeurbanne, Lyon, France
\item \Idef{org9}Centro de Aplicaciones Tecnol\'{o}gicas y Desarrollo Nuclear (CEADEN), Havana, Cuba
\item \Idef{org10}Centro de Investigaci\'{o}n y de Estudios Avanzados (CINVESTAV), Mexico City and M\'{e}rida, Mexico
\item \Idef{org11}Centro Fermi - Museo Storico della Fisica e Centro Studi e Ricerche ``Enrico Fermi', Rome, Italy
\item \Idef{org12}Chicago State University, Chicago, Illinois, United States
\item \Idef{org13}China Institute of Atomic Energy, Beijing, China
\item \Idef{org14}Chonbuk National University, Jeonju, Republic of Korea
\item \Idef{org15}Comenius University Bratislava, Faculty of Mathematics, Physics and Informatics, Bratislava, Slovakia
\item \Idef{org16}COMSATS Institute of Information Technology (CIIT), Islamabad, Pakistan
\item \Idef{org17}Creighton University, Omaha, Nebraska, United States
\item \Idef{org18}Department of Physics, Aligarh Muslim University, Aligarh, India
\item \Idef{org19}Department of Physics, Ohio State University, Columbus, Ohio, United States
\item \Idef{org20}Department of Physics, Pusan National University, Pusan, Republic of Korea
\item \Idef{org21}Department of Physics, Sejong University, Seoul, Republic of Korea
\item \Idef{org22}Department of Physics, University of California, Berkeley, California, United States
\item \Idef{org23}Department of Physics, University of Oslo, Oslo, Norway
\item \Idef{org24}Department of Physics and Technology, University of Bergen, Bergen, Norway
\item \Idef{org25}Dipartimento di Fisica dell'Universit\`{a} 'La Sapienza' and Sezione INFN, Rome, Italy
\item \Idef{org26}Dipartimento di Fisica dell'Universit\`{a} and Sezione INFN, Cagliari, Italy
\item \Idef{org27}Dipartimento di Fisica dell'Universit\`{a} and Sezione INFN, Trieste, Italy
\item \Idef{org28}Dipartimento di Fisica dell'Universit\`{a} and Sezione INFN, Turin, Italy
\item \Idef{org29}Dipartimento di Fisica e Astronomia dell'Universit\`{a} and Sezione INFN, Bologna, Italy
\item \Idef{org30}Dipartimento di Fisica e Astronomia dell'Universit\`{a} and Sezione INFN, Catania, Italy
\item \Idef{org31}Dipartimento di Fisica e Astronomia dell'Universit\`{a} and Sezione INFN, Padova, Italy
\item \Idef{org32}Dipartimento di Fisica `E.R.~Caianiello' dell'Universit\`{a} and Gruppo Collegato INFN, Salerno, Italy
\item \Idef{org33}Dipartimento DISAT del Politecnico and Sezione INFN, Turin, Italy
\item \Idef{org34}Dipartimento di Scienze e Innovazione Tecnologica dell'Universit\`{a} del Piemonte Orientale and INFN Sezione di Torino, Alessandria, Italy
\item \Idef{org35}Dipartimento Interateneo di Fisica `M.~Merlin' and Sezione INFN, Bari, Italy
\item \Idef{org36}European Organization for Nuclear Research (CERN), Geneva, Switzerland
\item \Idef{org37}Faculty of Engineering and Science, Western Norway University of Applied Sciences, Bergen, Norway
\item \Idef{org38}Faculty of Nuclear Sciences and Physical Engineering, Czech Technical University in Prague, Prague, Czech Republic
\item \Idef{org39}Faculty of Science, P.J.~\v{S}af\'{a}rik University, Ko\v{s}ice, Slovakia
\item \Idef{org40}Frankfurt Institute for Advanced Studies, Johann Wolfgang Goethe-Universit\"{a}t Frankfurt, Frankfurt, Germany
\item \Idef{org41}Gangneung-Wonju National University, Gangneung, Republic of Korea
\item \Idef{org42}Gauhati University, Department of Physics, Guwahati, India
\item \Idef{org43}Helmholtz-Institut f\"{u}r Strahlen- und Kernphysik, Rheinische Friedrich-Wilhelms-Universit\"{a}t Bonn, Bonn, Germany
\item \Idef{org44}Helsinki Institute of Physics (HIP), Helsinki, Finland
\item \Idef{org45}Hiroshima University, Hiroshima, Japan
\item \Idef{org46}Hochschule Worms, Zentrum  f\"{u}r Technologietransfer und Telekommunikation (ZTT), Worms, Germany
\item \Idef{org47}Horia Hulubei National Institute of Physics and Nuclear Engineering, Bucharest, Romania
\item \Idef{org48}Indian Institute of Technology Bombay (IIT), Mumbai, India
\item \Idef{org49}Indian Institute of Technology Indore, Indore, India
\item \Idef{org50}Indonesian Institute of Sciences, Jakarta, Indonesia
\item \Idef{org51}INFN, Laboratori Nazionali di Frascati, Frascati, Italy
\item \Idef{org52}INFN, Sezione di Bari, Bari, Italy
\item \Idef{org53}INFN, Sezione di Bologna, Bologna, Italy
\item \Idef{org54}INFN, Sezione di Cagliari, Cagliari, Italy
\item \Idef{org55}INFN, Sezione di Catania, Catania, Italy
\item \Idef{org56}INFN, Sezione di Padova, Padova, Italy
\item \Idef{org57}INFN, Sezione di Roma, Rome, Italy
\item \Idef{org58}INFN, Sezione di Torino, Turin, Italy
\item \Idef{org59}INFN, Sezione di Trieste, Trieste, Italy
\item \Idef{org60}Inha University, Incheon, Republic of Korea
\item \Idef{org61}Institut de Physique Nucl\'{e}aire d'Orsay (IPNO), Institut National de Physique Nucl\'{e}aire et de Physique des Particules (IN2P3/CNRS), Universit\'{e} de Paris-Sud, Universit\'{e} Paris-Saclay, Orsay, France
\item \Idef{org62}Institute for Nuclear Research, Academy of Sciences, Moscow, Russia
\item \Idef{org63}Institute for Subatomic Physics, Utrecht University/Nikhef, Utrecht, Netherlands
\item \Idef{org64}Institute for Theoretical and Experimental Physics, Moscow, Russia
\item \Idef{org65}Institute of Experimental Physics, Slovak Academy of Sciences, Ko\v{s}ice, Slovakia
\item \Idef{org66}Institute of Physics, Bhubaneswar, India
\item \Idef{org67}Institute of Physics of the Czech Academy of Sciences, Prague, Czech Republic
\item \Idef{org68}Institute of Space Science (ISS), Bucharest, Romania
\item \Idef{org69}Institut f\"{u}r Kernphysik, Johann Wolfgang Goethe-Universit\"{a}t Frankfurt, Frankfurt, Germany
\item \Idef{org70}Instituto de Ciencias Nucleares, Universidad Nacional Aut\'{o}noma de M\'{e}xico, Mexico City, Mexico
\item \Idef{org71}Instituto de F\'{i}sica, Universidade Federal do Rio Grande do Sul (UFRGS), Porto Alegre, Brazil
\item \Idef{org72}Instituto de F\'{\i}sica, Universidad Nacional Aut\'{o}noma de M\'{e}xico, Mexico City, Mexico
\item \Idef{org73}iThemba LABS, National Research Foundation, Somerset West, South Africa
\item \Idef{org74}Johann-Wolfgang-Goethe Universit\"{a}t Frankfurt Institut f\"{u}r Informatik, Fachbereich Informatik und Mathematik, Frankfurt, Germany
\item \Idef{org75}Joint Institute for Nuclear Research (JINR), Dubna, Russia
\item \Idef{org76}Konkuk University, Seoul, Republic of Korea
\item \Idef{org77}Korea Institute of Science and Technology Information, Daejeon, Republic of Korea
\item \Idef{org78}KTO Karatay University, Konya, Turkey
\item \Idef{org79}Laboratoire de Physique Subatomique et de Cosmologie, Universit\'{e} Grenoble-Alpes, CNRS-IN2P3, Grenoble, France
\item \Idef{org80}Lawrence Berkeley National Laboratory, Berkeley, California, United States
\item \Idef{org81}Lund University Department of Physics, Division of Particle Physics, Lund, Sweden
\item \Idef{org82}Nagasaki Institute of Applied Science, Nagasaki, Japan
\item \Idef{org83}Nara Women{'}s University (NWU), Nara, Japan
\item \Idef{org84}National and Kapodistrian University of Athens, School of Science, Department of Physics , Athens, Greece
\item \Idef{org85}National Centre for Nuclear Research, Warsaw, Poland
\item \Idef{org86}National Institute of Science Education and Research, HBNI, Jatni, India
\item \Idef{org87}National Nuclear Research Center, Baku, Azerbaijan
\item \Idef{org88}National Research Centre Kurchatov Institute, Moscow, Russia
\item \Idef{org89}Niels Bohr Institute, University of Copenhagen, Copenhagen, Denmark
\item \Idef{org90}Nikhef, National institute for subatomic physics, Amsterdam, Netherlands
\item \Idef{org91}NRC ¿Kurchatov Institute¿ ¿ IHEP , Protvino, Russia
\item \Idef{org92}NRNU Moscow Engineering Physics Institute, Moscow, Russia
\item \Idef{org93}Nuclear Physics Group, STFC Daresbury Laboratory, Daresbury, United Kingdom
\item \Idef{org94}Nuclear Physics Institute of the Czech Academy of Sciences, \v{R}e\v{z} u Prahy, Czech Republic
\item \Idef{org95}Oak Ridge National Laboratory, Oak Ridge, Tennessee, United States
\item \Idef{org96}Petersburg Nuclear Physics Institute, Gatchina, Russia
\item \Idef{org97}Physics department, Faculty of science, University of Zagreb, Zagreb, Croatia
\item \Idef{org98}Physics Department, Panjab University, Chandigarh, India
\item \Idef{org99}Physics Department, University of Jammu, Jammu, India
\item \Idef{org100}Physics Department, University of Rajasthan, Jaipur, India
\item \Idef{org101}Physikalisches Institut, Eberhard-Karls-Universit\"{a}t T\"{u}bingen, T\"{u}bingen, Germany
\item \Idef{org102}Physikalisches Institut, Ruprecht-Karls-Universit\"{a}t Heidelberg, Heidelberg, Germany
\item \Idef{org103}Physik Department, Technische Universit\"{a}t M\"{u}nchen, Munich, Germany
\item \Idef{org104}Research Division and ExtreMe Matter Institute EMMI, GSI Helmholtzzentrum f\"ur Schwerionenforschung GmbH, Darmstadt, Germany
\item \Idef{org105}Rudjer Bo\v{s}kovi\'{c} Institute, Zagreb, Croatia
\item \Idef{org106}Russian Federal Nuclear Center (VNIIEF), Sarov, Russia
\item \Idef{org107}Saha Institute of Nuclear Physics, Kolkata, India
\item \Idef{org108}School of Physics and Astronomy, University of Birmingham, Birmingham, United Kingdom
\item \Idef{org109}Secci\'{o}n F\'{\i}sica, Departamento de Ciencias, Pontificia Universidad Cat\'{o}lica del Per\'{u}, Lima, Peru
\item \Idef{org110}Shanghai Institute of Applied Physics, Shanghai, China
\item \Idef{org111}Stefan Meyer Institut f\"{u}r Subatomare Physik (SMI), Vienna, Austria
\item \Idef{org112}SUBATECH, IMT Atlantique, Universit\'{e} de Nantes, CNRS-IN2P3, Nantes, France
\item \Idef{org113}Suranaree University of Technology, Nakhon Ratchasima, Thailand
\item \Idef{org114}Technical University of Ko\v{s}ice, Ko\v{s}ice, Slovakia
\item \Idef{org115}Technische Universit\"{a}t M\"{u}nchen, Excellence Cluster 'Universe', Munich, Germany
\item \Idef{org116}The Henryk Niewodniczanski Institute of Nuclear Physics, Polish Academy of Sciences, Cracow, Poland
\item \Idef{org117}The University of Texas at Austin, Austin, Texas, United States
\item \Idef{org118}Universidad Aut\'{o}noma de Sinaloa, Culiac\'{a}n, Mexico
\item \Idef{org119}Universidade de S\~{a}o Paulo (USP), S\~{a}o Paulo, Brazil
\item \Idef{org120}Universidade Estadual de Campinas (UNICAMP), Campinas, Brazil
\item \Idef{org121}Universidade Federal do ABC, Santo Andre, Brazil
\item \Idef{org122}University College of Southeast Norway, Tonsberg, Norway
\item \Idef{org123}University of Cape Town, Cape Town, South Africa
\item \Idef{org124}University of Houston, Houston, Texas, United States
\item \Idef{org125}University of Jyv\"{a}skyl\"{a}, Jyv\"{a}skyl\"{a}, Finland
\item \Idef{org126}University of Liverpool, Liverpool, United Kingdom
\item \Idef{org127}University of Split, Faculty of Electrical Engineering, Mechanical Engineering and Naval Architecture, Split, Croatia
\item \Idef{org128}University of Tennessee, Knoxville, Tennessee, United States
\item \Idef{org129}University of the Witwatersrand, Johannesburg, South Africa
\item \Idef{org130}University of Tokyo, Tokyo, Japan
\item \Idef{org131}University of Tsukuba, Tsukuba, Japan
\item \Idef{org132}Universit\'{e} Clermont Auvergne, CNRS/IN2P3, LPC, Clermont-Ferrand, France
\item \Idef{org133}Universit\'{e} de Lyon, Universit\'{e} Lyon 1, CNRS/IN2P3, IPN-Lyon, Villeurbanne, Lyon, France
\item \Idef{org134}Universit\'{e} de Strasbourg, CNRS, IPHC UMR 7178, F-67000 Strasbourg, France, Strasbourg, France
\item \Idef{org135} Universit\'{e} Paris-Saclay Centre d¿\'Etudes de Saclay (CEA), IRFU, Department de Physique Nucl\'{e}aire (DPhN), Saclay, France
\item \Idef{org136}Universit\`{a} degli Studi di Pavia, Pavia, Italy
\item \Idef{org137}Universit\`{a} di Brescia, Brescia, Italy
\item \Idef{org138}V.~Fock Institute for Physics, St. Petersburg State University, St. Petersburg, Russia
\item \Idef{org139}Variable Energy Cyclotron Centre, Kolkata, India
\item \Idef{org140}Warsaw University of Technology, Warsaw, Poland
\item \Idef{org141}Wayne State University, Detroit, Michigan, United States
\item \Idef{org142}Westf\"{a}lische Wilhelms-Universit\"{a}t M\"{u}nster, Institut f\"{u}r Kernphysik, M\"{u}nster, Germany
\item \Idef{org143}Wigner Research Centre for Physics, Hungarian Academy of Sciences, Budapest, Hungary
\item \Idef{org144}Yale University, New Haven, Connecticut, United States
\item \Idef{org145}Yonsei University, Seoul, Republic of Korea
\end{Authlist}
\endgroup
\end{document}